\documentclass[useAMS,usenatbib]{mn2e}

\usepackage{txfonts,graphicx,amssymb,rotating}

\title[Dust dynamics during protoplanetary disc clearing]{Dust dynamics during protoplanetary disc clearing}

\author[R.D.~Alexander \& P.J.~Armitage]
  {R.D.~Alexander$^{1,}$\thanks{email: rda@jilau1.colorado.edu}
  and P.J.Armitage$^{1,2}$ \\
   $^1$ JILA, University of Colorado, Boulder, CO 80309-0440, USA\\
   $^2$ Department of Astrophysical and Planetary Sciences, University of Colorado, Boulder, CO 80309-0391, USA}

\begin{document}

\pagerange{\pageref{firstpage}--\pageref{lastpage}} \pubyear{2006}

\maketitle

\label{firstpage}

\begin{abstract}
We consider the dynamics of dust and gas during the clearing of protoplanetary discs.  We work within the context of a photoevaporation/viscous model for the evolution of the gas disc, and use a two-fluid model to study the dynamics of dust grains as the gas disc is cleared.  Small ($\lesssim 10$$\mu$m) grains remain well-coupled to the gas, but larger ($\sim1$mm) grains are subject to inward migration from large radii ($\sim50$AU), suggesting that the time-scale for grain growth in the outer disc is $\sim10^4$--$10^5$yr.  We describe in detail the observable appearance of discs during clearing, and find that pressure gradients in the gas disc result in a strong enhancement of the local dust-to-gas ratio in a ring near to the inner disc edge.  Lastly, we consider a simple model of the disc-planet interaction, and suggest that observations of disc masses and accretion rates provide a straightforward means of discriminating between different models of disc clearing.
\end{abstract}

\begin{keywords}
accretion, accretion discs -- circumstellar matter -- stars: pre-main-sequence -- planetary systems: protoplanetary discs -- planetary systems: formation
\end{keywords}


\section{Introduction}
It is now well accepted that newly-formed, solar-type stars are surrounded by primordial discs of gas and dust.  Such discs can occasionally be resolved directly \citep[e.g.][]{mod96}, but their presence is more commonly inferred from secondary observational diagnostics.  The most common such diagnostics are infrared excess, arising from the reprocessing of stellar radiation by warm dust in the inner disc disc \citep*{kh95,haisch01,hartmann05}, accretion signatures such as line emission and ultraviolet excess \citep*{heg95,cg98,gul00,muz00}, and millimetre continuum emission from cold dust in the outer disc \citep{beckwith90,aw05}.  These different diagnostics are usually seen together, and few transition objects are known to exist between the disc-bearing classical T Tauri stars (CTTs) and the disc-less weak-lined T Tauri stars (WTTs).  Since CTTs evolve into WTTs, the apparent lack of transition objects implies that the transition between the CTT and WTT states is very rapid \citep{skrutskie90,kh95,sp95,ww96}: typical estimates suggest that the discs are cleared on a time-scale of order $10^5$yr after a lifetime of a few Myr.  Moreover, the data show that the inner and outer discs are cleared almost simultaneously \citep[e.g.][]{aw05}, suggesting that the mechanism responsible for the rapid disc clearing operates over the entire radial extent of the disc, from $<0.1$AU to $>100$AU.

More recent studies have begun to provide some insight into the disc clearing process, and a handful of ``transitional'' discs, with properties between the CTT and WTT states, have now been well-studied (e.g.~GM Aur, \citealt{rice03,calvet05}; TW Hya, \citealt*{calvet02,eisner06}; CoKu Tau/4, \citealt{forrest04,dalessio05}).  These objects are characterised by a lack of near-infrared excess emission, but ``normal'' disc spectral energy distributions (SEDs) at wavelengths longer than $\sim10$$\mu$m.  However, some of these objects, such as GM Aur, are actively accreting and possess massive ($\sim 0.1$M$_{\odot}$) discs, while others, such as CoKu Tau/4, show no evidence of accretion and possess much less massive ($\sim 1$M$_{\mathrm {Jup}}$) discs.  Recently, \citet{sic06} identified 17 ``transition objects'', and found that roughly half showed evidence of accretion, while half did not.  It does not, therefore, seem likely that ``transition objects'' identified in this way (i.e.~on the basis of their SEDs) represent a homogenous class of objects.  However, all such objects do seem consistent with some form of ``inside-out'' disc clearing, during which the disc appears to have a large central cavity or hole, at least in the dust disc \citep[see also][]{padgett06}.  The mechanism(s) responsible for producing such holes remain uncertain, however, and previous studies have variously invoked the presence of a planet \citep[e.g.][]{rice03,quillen04}, dust evolution \citep[e.g.][]{wilner05} or photoevaporation \citep[e.g.][]{acp06b,goto06} to explain the observed ``holes''.

In this paper we seek to explore the behaviour of the dust component of the disc during the clearing phase, within the context of photoevaporation models of disc evolution.  We also seek observable diagnostics that can be used to discriminate between different models of disc clearing.  Photoevaporation occurs when ultraviolet radiation heats the disc surface sufficiently that it becomes unbound from the central star and flows from the disc as a wind \citep[see the reviews by][and references within]{holl_ppiv,dull_ppv}.  Here we focus on the effect of ionizing radiation from the central star, as the effects of external photoevaporation from nearby O-stars are only significant for the minority of TTs which are found near to the centre of massive clusters \citep{sh99,ec06}.  It has been demonstrated that it is reasonable to treat the central star as having an ionizing luminosity of order $\Phi = 10^{42}$photons s$^{-1}$, presumably arising in the stellar chromosphere, which is approximately constant over the duration of the TT phase \citep*{chrom}.  Most previous studies of the evolution of photoevaporating discs \citep*[e.g.][]{cc01,mjh03,acp03,acp06b} have modelled the evolution of the gas disc only, and simply assumed a constant dust-to-gas ratio in order to compare to observed data.  These studies have shown that models which incorporate photoevaporation and viscous accretion can reproduce a number of observed properties, such as the rapid transition from the CTT to WTT state and the near-simultaneous clearing of the inner and outer discs.  The basic premise of such models is that the low mass-loss rate of the wind ($\sim 10^{-10}$M$_{\odot}$yr$^{-1}$) is initially negligible compared to the disc accretion rate.  However, the disc accretion rate falls with time, eventually becoming comparable to the (constant) wind rate.  At this point the wind opens a gap in the disc, depriving the inner disc of resupply.  The inner disc drains on to the star on a viscous time-scale, and from this point onwards the wind dominates the evolution, resulting in a rapid clearing of the disc from the inside-out.  

\citet*[][see also \citealt{tl05}]{tcl05} combined a viscous/photoevaporation model for the evolution of gas discs with a two-fluid model for the evolution of the dust, and studied the evolution of mm-size grains over the lifetime of the disc.  They found that mm grains migrate inwards (due to headwind drag) on fairly short time-scales, typically $\sim 1$Myr or less.  Consequently, in their model the mm-size grains were removed from TT discs prior to the dispersal of the gas disc, and they predicted the existence of a population of dust-poor gas discs.  More recently, however, \citet{aw05} surveyed more than 100 TTs in the Taurus-Auriga cloud and found that essentially all objects ($>90$\%) with optically thick, accreting inner discs also show a strong millimetre flux.  In addition, only a handful of objects ($<10$\%) with no infrared excess or accretion signatures were detected at millimetre wavelengths.  This tells us, in contrast with the predictions of \citet{tcl05}, that mm-sized dust in the outer disc is dispersed at the same time as the gas disc.  Moreover, this suggests that mm grains in the outer disc must be replenished throughout the lifetime of the CTT phase, as the inward migration time-scale of these grains is much shorter than the typical CTT lifetime.  In addition, detailed models of dust coagulation have shown that replenishment of grains is necessary even to sustain a population of small, micron-sized grains, over the (several Myr) duration of the CTT phase \citep{dd05}.

Consequently, here we make the assumption that the dust-to-gas ratio (for all grain sizes) is constant at the beginning of the clearing phase, and consider the differential motion of the gas and dust during the photoevaporatively-induced transition.  Essentially we assume that some replenishment process(es), such as that suggested by \citet{dd05}, maintains a quasi-equilibrium in the grain population up to the point where photoevaporation alters the gas dynamics significantly, and we then follow the differential evolution of the dust and gas discs through the transitional phase.  We use a model similar to that presented by \citet{tcl05}, but also include updated photoevaporation models \citep{font04,acp06a,acp06b} which radically alter the evolution of the outer gas disc at late times.  In addition we consider a range of grain sizes, from small $\mu$m grains up to larger mm-sized particles.  The structure of the paper is as follows.  In Section \ref{sec:model} we describe our disc model, and present the results in Section \ref{sec:res}.  We discuss the implications of these results, and the limitations of our analysis, in Section \ref{sec:dis}.  In Section \ref{sec:holes} we compare our results to those from a simple model of the disc-planet interaction, and suggest a simple observational diagnostic that can be used to discriminate between these different models, before summarizing our conclusions in Section \ref{sec:summary}.


\section{Model}\label{sec:model}
In order to model the dynamics of dust and gas during the transition phase, we adopt a model similar to that used by \citet{tcl05}.  We assume that the gaseous component of the disc evolves subject to viscosity and photoevaporation, and that the dust component is subject to diffusion and gas-drag forces.  We neglect other effects, such as grain replenishment, radiation pressure, or Poynting-Robertson drag; the possible consequences of these simplifications are discussed in Section \ref{sec:dis}.

\subsection{Gas disc}\label{sec:gas}
The evolution of the gas surface density, $\Sigma_{\mathrm g}(R,t)$ is governed by the diffusion equation \citep{lbp74,pringle81}
\begin{equation}\label{eq:1ddiff}\label{eq:gas_evo}
\frac{\partial \Sigma_{\mathrm g}}{\partial t} = \frac{3}{R}\frac{\partial}{\partial R}\left[ R^{1/2} \frac{\partial}{\partial R}\left(\nu \Sigma_{\mathrm g} R^{1/2}\right) \right] - \dot{\Sigma}_{\mathrm {wind}}(R,t) \, ,
\end{equation}
where $t$ is time, $R$ is cylindrical radius, $\nu$ is the kinematic viscosity and the term $\dot{\Sigma}_{\mathrm {wind}}(R,t)$ represents the mass-loss due to photoevaporation.    

\subsubsection{Viscosity}\label{sec:visc}
The evolution of the gas disc is determined primarily by the form of the kinematic viscosity $\nu$, which governs the transport of angular momentum in the disc.  The details of how angular momentum is transported in accretion discs are still subject to much debate \citep[e.g.][]{gammie96,bh98}, but for simplicity we adopt an alpha-disc model \citep{ss73}.  The viscosity takes the form
\begin{equation}
\nu(R) = \alpha \Omega H^2 \, ,
\end{equation}
where $\alpha$ is the standard \citet{ss73} viscosity parameter, $\Omega(R) = \sqrt{GM_*/R^3}$ is the orbital frequency, and $H(R)$ is the disc scale-height.  We adopt a scale-height consistent with a ``flared disc'' model \citep[e.g.][]{kh87}, which takes the form
\begin{equation}
H(R) \propto R^q \, ,
\end{equation}
with the power-law index $q=5/4$.  We normalise this relationship so that the disc aspect ratio $H/R=0.0333$ at $R=1$AU.  (With this scaling we therefore have $H/R=0.019$ at $R=0.1$AU, and $H/R=0.105$ at $R=100$AU.)  This choice of the index $q$ implies that the disc midplane temperature scales as $R^{-1/2}$, and results in a viscosity which scales linearly with radius.  Arguments in favour of such a viscosity law were discussed by \citet{hcga98}, and recent observations also support this parametrization \citep{aw06}.  We consider two viscosity models within this framework: a ``high-viscosity'' model with $\alpha=0.01$, and a ``low-viscosity'' model with $\alpha=0.001$.

\subsubsection{Wind model}\label{sec:wind}
Here we seek to model the transitional phase more accurately than before, and so the wind term is given by
\begin{equation}
\dot{\Sigma}_{\mathrm {wind}}(R,t) = \dot{\Sigma}_{\mathrm {diffuse}} + \dot{\Sigma}_{\mathrm {direct}} \, ,
\end{equation}
where
$\dot{\Sigma}_{\mathrm {diffuse}}$ and $\dot{\Sigma}_{\mathrm {direct}}$ represent the contributions to the wind from the diffuse and direct stellar radiation fields respectively.  We adopt the diffuse wind profile from \citet{font04}, and the direct wind profile from \citet{acp06a} (with $H/R=0.05$).  The numerical forms of these wind profiles are defined in Appendix \ref{sec:wind_profs}.  The direct profile is defined in \citet{acp06a} as a power-law: it is only defined for $R>R_{\mathrm {in}}$, and diverges to small radii.  In order to avoid numerical problems we smooth this profile at the inner disc edge $R_{\mathrm {in}}$ by multiplying $\dot{\Sigma}_{\mathrm {direct}}(R)$ by the smoothing function
\begin{equation}\label{eq:smooth}
f(R) = \left[ 1+ \exp\left(-\frac{R-R_{\mathrm {in}}}{H_{\mathrm {in}}}\right) \right]^{-1} \, ,
\end{equation}
where $H_{\mathrm {in}}$ is the disc scale-height at the inner edge.  Both the diffuse and direct wind rates scale as the square root of the respective ionizing fluxes, and we model the transition by scaling these fluxes appropriately.  The diffuse flux $\Phi_{\mathrm {diffuse}}$ arises from recombinations of atomic hydrogen in the disc atmosphere, and was modelled in detail by \citet{holl94}.  Here we note that as the inner disc drains the structure of this atmosphere is modified, and the contribution from the atmosphere at a given radius is essentially removed if the disc at that radius is optically thin in the vertical direction to ionizing photons.  We denote this radius (at which the disc becomes optically thin in the vertical direction) by $R_{\mathrm {thin}}$ and, following \citet{holl94}, we assume that the diffuse ionizing flux reaching larger radii is dominated by the flux emitted at this radius.  This flux is subject to geometric dilution, and consequently we scale the diffuse ionizing flux as
\begin{equation}\label{eq:thickthin}
\Phi_{\mathrm {diffuse}} = \left\{ \begin{array}{ll}
\Phi \left(\frac{R_{\mathrm {thin}}}{R_{\mathrm {in}}}\right)^2 & \textrm{if } R_{\mathrm {thin}} < R_{\mathrm {crit}}\\
\Phi & \mathrm{otherwise}
\end{array} \right.
\end{equation}
where $\Phi$ is the stellar ionizing flux and $R_{\mathrm {in}}$ is the radius of the inner disc edge outside the gap.  (Note that $R_{\mathrm {thin}} \le R_{\mathrm {in}}$, by construction.)  $R_{\mathrm {crit}}$ is the ``critical radius'' at which the gap opens, which is given by \citep{liffman03,font04,dull_ppv}
\begin{equation}
R_{\mathrm {crit}} \simeq 1.4 \left( \frac{M_*}{1 \mathrm M_{\odot}} \right) \mathrm {AU} \, ,
\end{equation}
and is approximately one-fifth of the ``gravitational radius'' defined by \citet{holl94}.  If the surface density in the inner disc is sufficiently small that the disc is optically thin to ionizing photons $R_{\mathrm {thin}}$ is located where
\begin{equation}
\Sigma_{\mathrm g}(R_{\mathrm {thin}}) = m_{\mathrm H} \sigma_{13.6{\mathrm eV}}^{-1} \, .
\end{equation}
Here $\sigma_{13.6{\mathrm eV}} = 6.3\times10^{-18}$cm$^{2}$ is the absorption cross section for ionizing photons \citep{allen}, and $m_{\mathrm H}$ is the mass of a hydrogen atom.

The direct radiation field reaching the outer disc is simply the stellar field attenuated by material in the inner disc.  Consequently we evaluate $\Phi_{\mathrm {direct}}$ as
\begin{equation}
\Phi_{\mathrm {direct}} = \Phi \exp(-\tau) \, ,
\end{equation}
where $\tau = N\sigma_{13.6{\mathrm eV}}$ is the radial optical depth to ionizing photons due to attenuation by gas with radial column density $N$.  We neglect attenuation due to dust in the inner disc (see Section \ref{sec:gas_res}).  \citet{acp06a} found that most of the mass-loss due to the direct field comes from 1--3 scale-heights above the disc midplane, so we evaluate the gas column $N$ as
\begin{equation}
N = \int_0^{R_{\mathrm {in}}} n_{\mathrm g} dR \, ,
\end{equation}
where the gas number density $n_{\mathrm g}(R)$ is evaluated from the surface density as
\begin{equation}\label{eq:gas_den}
n_{\mathrm g}(R) = \frac{\Sigma_{\mathrm g}(R)}{\sqrt{2 \pi} H(R) m_{\mathrm H}} \exp(-\chi^2/2) \, .
\end{equation}
Here the exponential term accounts for evaluating the density at $\chi$ scale-heights above the midplane.  We adopt $\chi=2.5$, but note that the evolution is not especially sensitive to the choice of $\chi$.

Consequently the transition from the diffuse to direct regime is treated more accurately than in previous work \citep[e.g.][]{acp06b}.  At early times the inner disc is extremely optically thick to ionizing photons, so the wind profile is simply the ``diffuse'' profile of \citet{font04}.  Similarly at late times the inner disc is optically thin to ionizing photons at all radii, and the wind is simply the ``direct'' profile of \citet{acp06a}.  A more detailed approach could consider the progression of the ionization front through the (draining) inner disc, using something akin to a Str\"omgrem criterion.  Evaluating the attenuation in this manner results in less attenuation than the parametrization described above, and leads to a slightly earlier rise in the direct field.  However, this form is also rather sensitive to the choice of the disc scale-height, and the difference between the two parametrizations is not significant.  (The direct field ``turns on'' slightly earlier, but the difference is only a small fraction of the already short viscous time-scale of the inner disc.)  Consequently we adopt the more conservative parametrization described in Equations \ref{eq:thickthin}--\ref{eq:gas_den}.

\subsubsection{Initial conditions}\label{sec:ic}
We are only interested in the dynamics during the transition from CTT to WTT, so we choose initial conditions that reflect this fact.  Previous photoevaporation models \citep[e.g.][]{cc01,acp06b} have shown that the wind starts to influence the evolution when the disc accretion rate drops to around 1/4 of the integrated (diffuse-field) wind rate, so we choose initial conditions to mimic this stage in the evolution of the gas disc.  We adopt an ionizing flux of $\Phi=10^{42}$photon s$^{-1}$ throughout.  This results in a (diffuse-field) wind mass-loss rate of $\simeq 4\times10^{-10}$M$_{\odot}$yr$^{-1}$, so we choose an initial accretion rate of $\dot{M}_0 = 1.0\times10^{-10}$M$_{\odot}$yr$^{-1}$.  We define the initial surface density profile to be that of a steady disc \citep[e.g.][]{pringle81}:
\begin{equation}
\Sigma_{\mathrm g}(R) = \frac{\dot{M}_0}{3 \pi \nu(R)}\left(1-\sqrt{\frac{R_{\mathrm {in}}}{R}}\right)\exp\left(-\frac{R}{R_{\mathrm d}}\right) \, .
\end{equation}
Here the second term is due to the presence of a (zero-torque) inner boundary at $R_{\mathrm {in}}$, and the exponential cutoff beyond radius $R_{\mathrm d}$ defines the size of the disc (and prevents divergence of the disc mass to large radii).  We adopt $R_{\mathrm d} = 500$AU throughout.

\subsection{Dust disc}\label{sec:dust}
In order to model the dynamics of the dust component of the disc, we follow the approach adopted by \citet[][see also \citealt{tl02}]{tl05}.  We make the simplifying assumption that the grains are not replenished over the short time-scales in which we are interested, so the only way that dust mass can be ``lost'' is by accretion on to the star.  The evolution of the dust surface density $\Sigma_{\mathrm d}(R,t)$ is governed by the continuity equation
\begin{equation}\label{eq:dust}
\frac{\partial \Sigma_{\mathrm d}}{\partial t} + \frac{1}{R}\frac{\partial}{\partial R} \left[ R(F_{\mathrm {dif}} + \Sigma_{\mathrm d} u_{\mathrm d}) \right] = 0 \, .
\end{equation}
Here $F_{\mathrm {dif}}$ is the diffusive mass flux, and $u_{\mathrm d}$ is the radial dust velocity induced by gas-drag.  Our initial conditions result in an outward pressure gradient in the gas disc, which causes the gas at a given radius to orbit at a slightly sub-Keplerian velocity.  This gives rise to a headwind drag force \citep{w77}, which removes angular momentum from the dust and results in inward radial migration of the grains.  The form of this velocity was studied in depth by \citet{tl02}, and we adopt their prescription for $u_{\mathrm d}$:
\begin{equation}\label{eq:v_adv}
u_{\mathrm d} = \frac{T_{\mathrm s}^{-1}u_{\mathrm g} - \eta u_{\mathrm K}}{T_{\mathrm s} + T_{\mathrm s}^{-1}} \, .
\end{equation}
Here $u_{\mathrm K} = (GM_*/R)^{1/2}$ is the Keplerian orbital velocity, and the gas radial velocity $u_{\mathrm g}$ is given by
\begin{equation}
u_{\mathrm g} = -\frac{3}{R^{1/2}\Sigma_{\mathrm g}} \frac{d}{dR}\left(R^{1/2} \nu \Sigma_{\mathrm g} \right) \, .
\end{equation}
$T_{\mathrm s}$ is the non-dimensional stopping time due to gas-drag on grains at the disc midplane (i.e.~the stopping time divided by the orbital time), and is given by
\begin{equation}
T_{\mathrm s} = \frac{\pi \rho_{\mathrm d} s}{2 \Sigma_{\mathrm g}}
\end{equation}
for grains of radius $s$ and density $\rho_{\mathrm d}$.  We fix the density to have the fiducial value $\rho_{\mathrm d} = 1$g cm$^{-3}$, and explore various values of $s$.  $\eta$ is defined as as the ratio of the pressure gradient to the gravity, and is given by\footnote{Note that our definition of the power-law index $q$ differs from that adopted by \citet{tl05}; Equation \ref{eq:eta} has been adjusted accordingly.}
\begin{equation}\label{eq:eta}
\eta = - \left(\frac{H}{R}\right)^2 \left(\frac{d \log \Sigma_{\mathrm g}}{d \log R} + (q-3) \right) \, .
\end{equation}

The diffusive term $F_{\mathrm {dif}}$ was derived by \citet{cp87}, for the case where $\Sigma_{\mathrm d}/\Sigma_{\mathrm g} \ll 1$, and in this case is given by
\begin{equation}
F_{\mathrm {dif}} = - D \Sigma_{\mathrm g} \frac{\partial}{\partial R} \left( \frac{\Sigma_{\mathrm d}}{\Sigma_{\mathrm g}} \right) \, .
\end{equation}
The relationship between the dust diffusion coefficient $D$ and the gas diffusion coefficient (viscosity) $\nu$ is not fully understood, with hydrodynamics models suggesting ratios of $\nu/D$ that range from $\simeq0.5$--10 \citep*{csp05,jk05,turner06,fp06}.  For simplicity we assume that the dust diffuses in a comparable manner to the gas, and therefore adopt $D=\nu$,   As long as the concentration $\Sigma_{\mathrm d}/\Sigma_{\mathrm g} \ll 1$, or as long as the advective term dominates the dust evolution, this form for the diffusive term is valid.  However, our model breaks down at late times, as the dust-to-gas ratio at the inner edge of the gas disc can become large.  Given the other uncertainties in the model, we make no attempt to correct for this and simply stop the calculations at this point.  This is addressed Section \ref{sec:edge}.  

The ability of the photoevaporative wind to carry dust grains depends on whether gas drag from the wind can overcome gravity and lift the grains away from the disc.  As shown by \citet{tcl05}, for spherical grains in a laminar flow the drag force from the wind is approximately
\begin{equation}
F_{\mathrm w} \simeq \frac{m_{\mathrm d} \rho_{\mathrm {w}} c_{\mathrm s}^2}{\rho_{\mathrm d} s} \, ,
\end{equation}
where $m_{\mathrm d}$ is the mass of a grain, $\rho_{\mathrm {w}}$ is the density of the gas at the base of the flow, and the flow is launched at the sound speed $c_{\mathrm s}=10$km s$^{-1}$.  The wind density peaks close to the inner edge of the disc in the direct field case \citep{acp06a}, and if we compare $F_{\mathrm w}$ to the weight of the grains, $F_{\mathrm g} = GM_* m_{\mathrm d}/R^2$, we see that that $F_{\mathrm w} < F_{\mathrm g}$ as long as $s \gtrsim 1$$\mu$m.  Moreover, we note that dust grains are expected to sediment towards the disc midplane as the disc evolves, but the photoevaporative wind typically flows from several scale-heights above the disc midplane.  Thus only small, sub-$\mu$m sized grains can be carried by the wind, and only then if something (such as turbulence) lifts them to the upper regions of the disc.  Consequently we neglect dust mass-loss due to the photoevaporative wind.

We define the initial conditions such that $\Sigma_{\mathrm d}/\Sigma_{\mathrm g} = 0.01$ at all radii.  Our model considers only a single grain size $s$, so formally this implies that a single grain size has a dust mass that is 1\% of the gas mass.  However, we note that the evolution of the dust disc in independent of the absolute value of $\Sigma_{\mathrm d}$, as long as the dust-to-gas ratio remains small.  

\subsection{Numerical method}\label{sec:numerics}
We solve the diffusion equation for the gas surface density (Equation \ref{eq:gas_evo}) numerically using a standard first-order explicit scheme \citep*[e.g.][]{pvw86}, on a grid of points equispaced in $R^{1/2}$.  We adopt zero-torque boundary conditions (i.e.~we set $\Sigma_{\mathrm g}=0$ in the boundary cells).  

The equation for the evolution of the dust is evaluated simultaneously on the same spatial grid.  We use a staggered grid, with scalar quantities evaluated at at zone centres and vectors (velocities) at zone faces.  The integration is explicit, using a first-order method for the diffusive term and a second-order (van Leer) method for the advective term.  Again we adopt ``outflow'' boundary conditions by setting $\Sigma_{\mathrm d} = 0$ in the boundary cells.  However, the inner boundary condition introduces an artificial inward pressure gradient in the gas, resulting in artificial outward advection of dust near the boundary, so we neglect the pressure gradient term in Equation \ref{eq:eta} in the region close to the inner boundary.  A similar, but physically real, inward pressure gradient is introduced by gaps in the gas disc which form as a result of photoevaporation.  However, the numerical form of the gap can result in rather large radial dust velocities, so in such cases we limit the magnitude of the (outward) dust velocity to be less than or equal to that of the (inward) gas velocity.

The inclusion of the photoevaporative wind at a stage in the calculation where the dust surface density is not negligible introduces an additional numerical complication compared with the calculations of \citet{tl05} or \citet{tcl05}, namely that the concentration term $\Sigma_{\mathrm d}/\Sigma_{\mathrm g}$ diverges as the gas surface density becomes small (at the gap in the gas disc).  In regions with no gas, the dust remains at a constant radius on a Keplerian orbit, so in cells where $\Sigma_{\mathrm g}=0$ we set $\partial \Sigma_{\mathrm d}/\partial t =0$.  However the divergence of the concentration term also causes a problem in cells adjacent to those where $\Sigma_{\mathrm g}=0$, as our evaluation of $\partial \Sigma_{\mathrm d}/\partial t $ uses ``three-cell'' derivatives for both the advective and diffusive terms.  In this case we revert to a first-order (donor cell) derivative for the advective term, and evaluate the diffusive term one cell further from the ``gap''.

This procedure successfully removes the troublesome divergent cells from the numerical calculation, but at the cost of reduced accuracy close to the edges of the gas disc.  The treatment of the diffusive term in particular causes problems, as the surface density of the gas rises sharply near the gap, and the accuracy is limited by the grid resolution.  Essentially we evaluate the diffusive term one cell-width away from where we should, so lowering the grid resolution reduces the accuracy still further.  The consequence of this inaccuracy is that dust mass is not exactly conserved close to the gap in the gas disc.  In order to minimize the effect this has on the overall accuracy of the calculation we take a ``brute force'' approach and simply use a large number of grid cells.  We adopt a grid spacing of $\Delta R^{1/2} = 0.0125$AU$^{1/2}$ and model the disc over the radial range [0.1AU,2000AU], so consequently we use 7108 grid cells.  At this resolution, dust mass is conserved to better than $\pm5$\% in all our models.


\section{Results}\label{sec:res}
\subsection{Gas disc}\label{sec:gas_res}
We first describe the evolution of the gas disc alone.  The evolution does not differ significantly from the models presented in \citet{acp06b}, and we refer the reader to this work for a detailed description of the gas disc evolution.  The evolution of the gas surface density $\Sigma_{\mathrm g}$ in the high- and low- viscosity models, can be seen in Figs.\ref{fig:dust_high} \& \ref{fig:dust_low} respectively.  Our models begin (i.e. they have $t=0$) at the point where the wind begins to influence the evolution, so the wind opens a gap in the disc rather rapidly, on the viscous time-scale of the inner disc.  Once the surface density in the inner disc drops below $\Sigma_{\mathrm g} \simeq 10^{-6}$g cm$^{-2}$, the inner disc becomes optically thin to ionizing photons and the direct field begins to dominate the evolution, rapidly clearing the outer disc.  

By comparing the two viscosity models, we see that the absolute clearing time-scale varies almost exactly as $1/\alpha$.  In the diffuse wind regime the evolution is driven by the viscosity, so the time-scale to open the gap scales exactly as $1/\alpha$.  Once the direct field takes over, however, the wind dominates the evolution (rather than the viscosity, see discussion in \citealt{acp06b}) and we see a small departure from this $1/\alpha$ scaling.  This occurs because the gap is opened at constant accretion rate, so the low-viscosity disc is more massive and the wind takes correspondingly longer to disperse it.  

In the high-viscosity model, with $\alpha=0.01$, the initial mass of the gas disc is $4.6\times10^{-4}$M$_{\odot}$, or around 0.48 Jupiter masses.  The effect of the wind on the disc is apparent almost immediately, and a well-defined gap in the disc appears at $t=3.3\times10^4$yr.  Cut-off from resupply, the inner disc then drains on its viscous time-scale.  In the low-viscosity case ($\alpha=0.001$), the initial disc mass is larger by a factor of 10 ($4.6\times10^{-3}$M$_{\odot}$, or around 4.8 Jupiter masses), and the evolution slower by a factor of 10 (i.e.~the gap opens at $t=3.3\times10^5$yr).  As the inner disc drains it eventually becomes optically thin to ionizing photons, and the direct field begins to clear the outer disc: this occurs at $t=4.1\times10^4$yr for $\alpha=0.01$, and $4.3\times10^5$yr for $\alpha=0.001$.  We stop the calculations at $t=5.5\times10^4$yr ($5.5\times10^5$yr for the low-viscosity model), at which point the inner edge of the disc has reached approximately 60.5AU (47.5AU).

The more accurate treatment of the diffuse-direct transition (described in Section \ref{sec:wind}) results in a smoother transition between the two regimes than seen by \citet{acp06b}, but the dynamics are not significantly modified.  As we show in the following Section, the dust in the inner disc is in fact removed more rapidly than the gas, so the inclusion of dust opacity has no effect on the results.  As mentioned above, a more accurate treatment still could consider the evolution of the ionization front through the draining inner disc, using something akin to a Str\"omgren criterion, but this should not result in significantly different behaviour\footnote{This is done by solving the recombination balance equation $\Phi = \int \alpha_{\mathrm B} n^2 dV$ to find the gas number density $n$, where $dV $ is the volume element and $\alpha_{\mathrm B}$ is the appropriate (Case B) recombination coefficient.}.  In this case the inner disc will become fully ionized when the surface density drops below $\simeq 10^{-4}$g cm$^{-2}$ (the exact value depends rather sensitively on the disc thickness, as the recombination rate scales with the square of the gas density), so we expect that the direct field will become significant once the total gas mass in the inner disc falls below $\sim 10^{-9}$M$_{\odot}$.  This results in a the direct field becoming significant slightly earlier than in our model, but the difference is not significant for the evolution of the disc.  Moreover, given the other uncertainties in this process, such as the behaviour of angular momentum transport in the low-surface density inner disc, we consider this treatment to be sufficiently accurate.

\subsection{Dust disc}\label{sec:dust_res}
\begin{figure}
\centering
        \resizebox{\hsize}{!}{
        \includegraphics[angle=270]{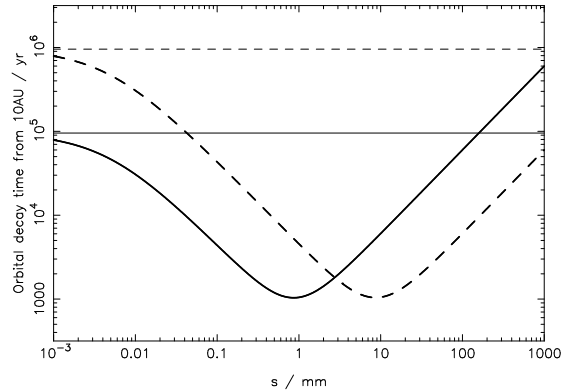}
        }
        \caption{Orbital decay times, defined as $t = -R/u$, from $R=10AU$ in the initial, steady, disc, plotted as a function of grain size $s$.  The heavy lines show the decay time due to advection, while the light lines show the (constant) viscous time-scale for diffusion of the gas.  The solid lines show the results for the high-viscosity model; the dashed lines the low-viscosity model.}
        \label{fig:timescales}
\end{figure}
In order to understand the evolution of the dust disc, it is instructive to re-cast Equation \ref{eq:dust} in terms of a diffusive radial dust velocity $u_{\mathrm {dif}}$, so that
\begin{equation}
\frac{\partial \Sigma_{\mathrm d}}{\partial t} + \frac{1}{R}\frac{\partial}{\partial R} \left[ R \Sigma_{\mathrm d} (u_{\mathrm {dif}} + u_{\mathrm d}) \right] = 0 \, .
\end{equation}
The advective velocity $u_{\mathrm d}$ is given in Equation \ref{eq:v_adv}, and depends on both the grain size and gas density, while we can express the diffusive velocity as
\begin{equation}
u_{\mathrm {dif}} = -\frac{D}{R} \frac{d \log (\Sigma_{\mathrm d}/\Sigma_{\mathrm g})}{d \log R} \, .
\end{equation}
The diffusive dust velocity depends on the viscosity, and also the dust concentration gradient.  $u_{\mathrm {dif}}$ is therefore comparable to the diffusive gas velocity $u_{\mathrm g}$, but is usually slightly larger because of the influence of the concentration gradient term.  The diffusive component acts to smooth out the concentration gradient, while the advective term always acts radially inward (until the gap in the gas disc opens).  We see, therefore, that the evolution of the dust grains is determined by competition between these two components; whichever acts more rapidly dominates the evolution.

\begin{figure*}
\centering
        \resizebox{\hsize}{!}{
        \includegraphics[angle=270]{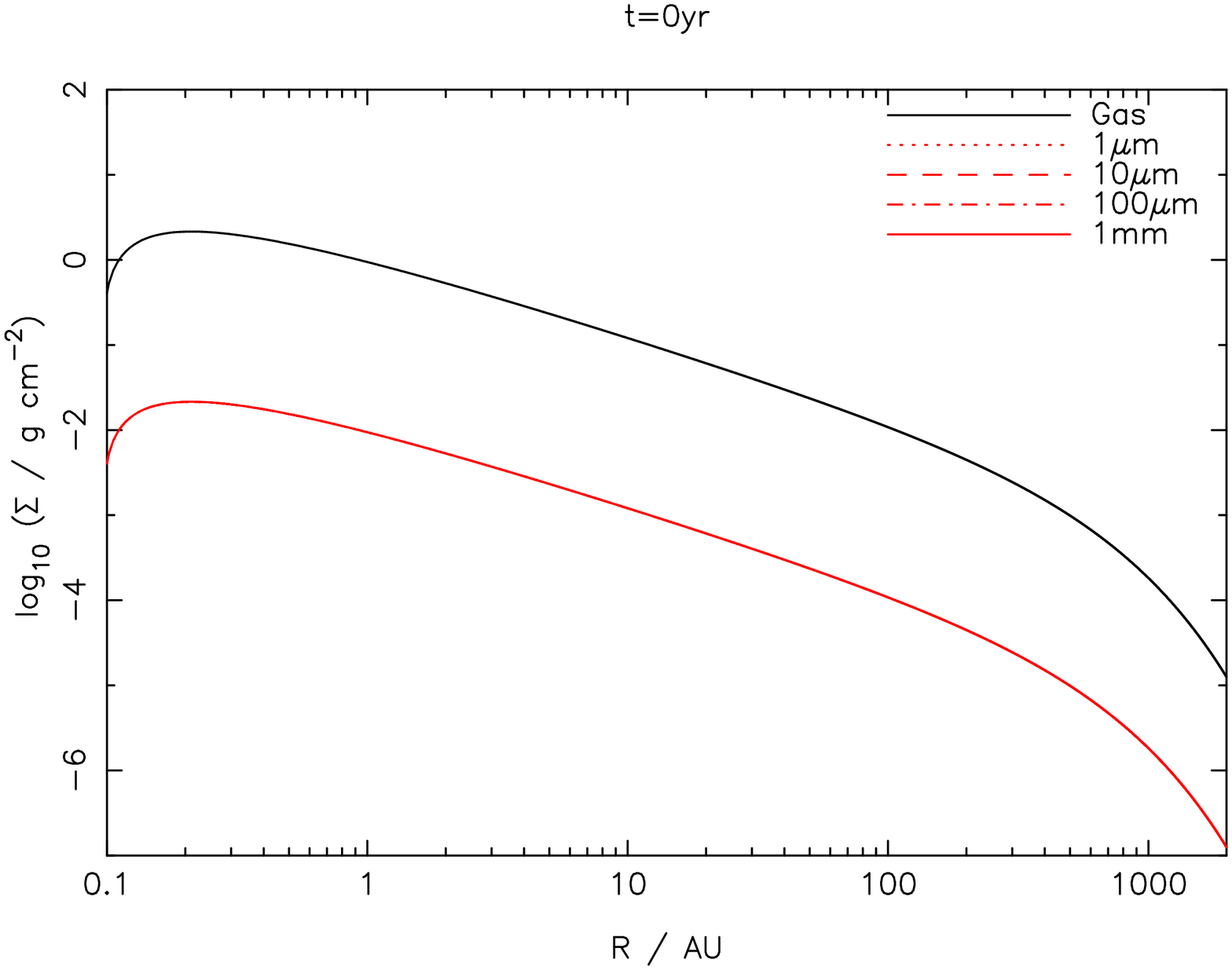}
        \includegraphics[angle=270]{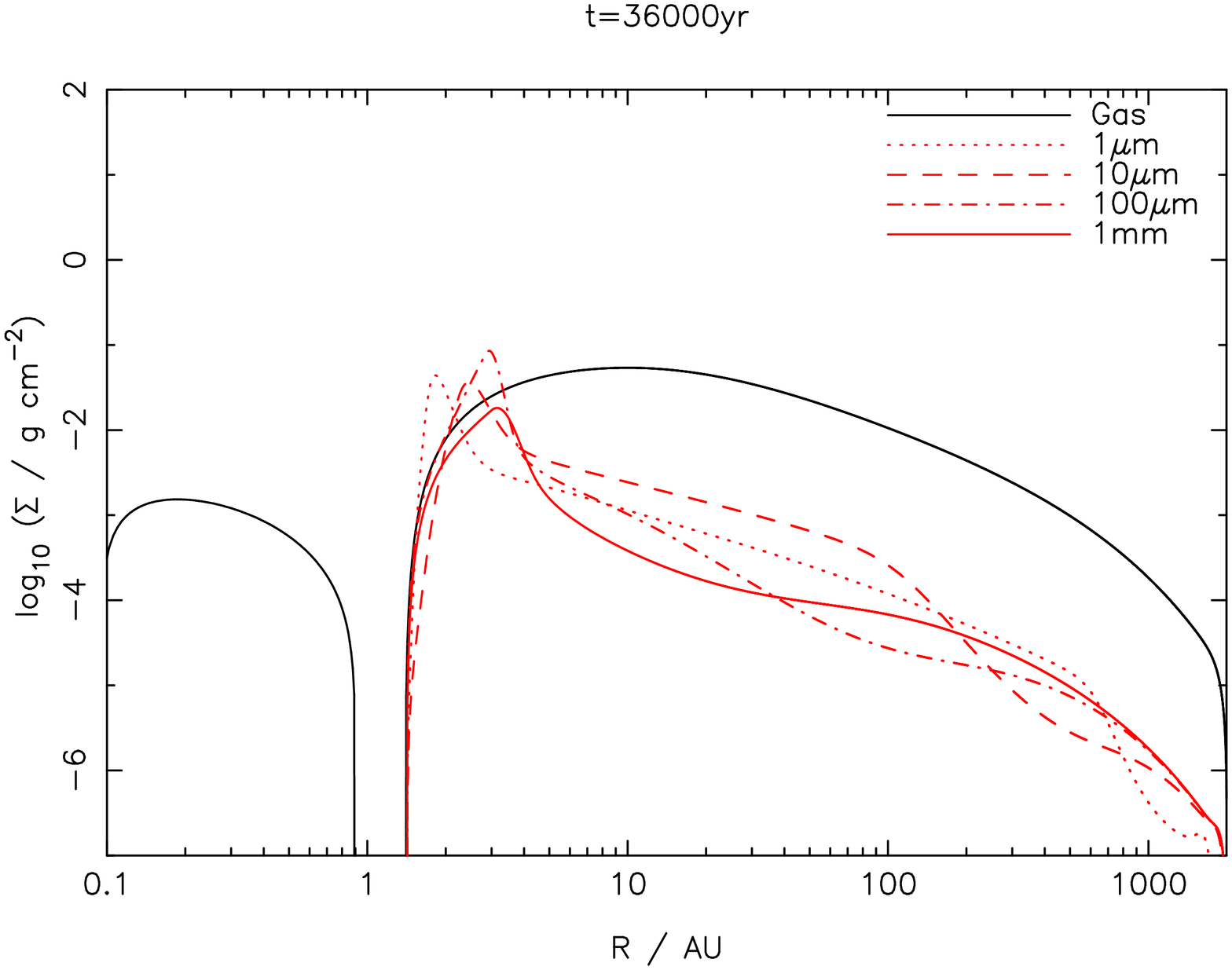}
        }
        \resizebox{\hsize}{!}{
        \includegraphics[angle=270]{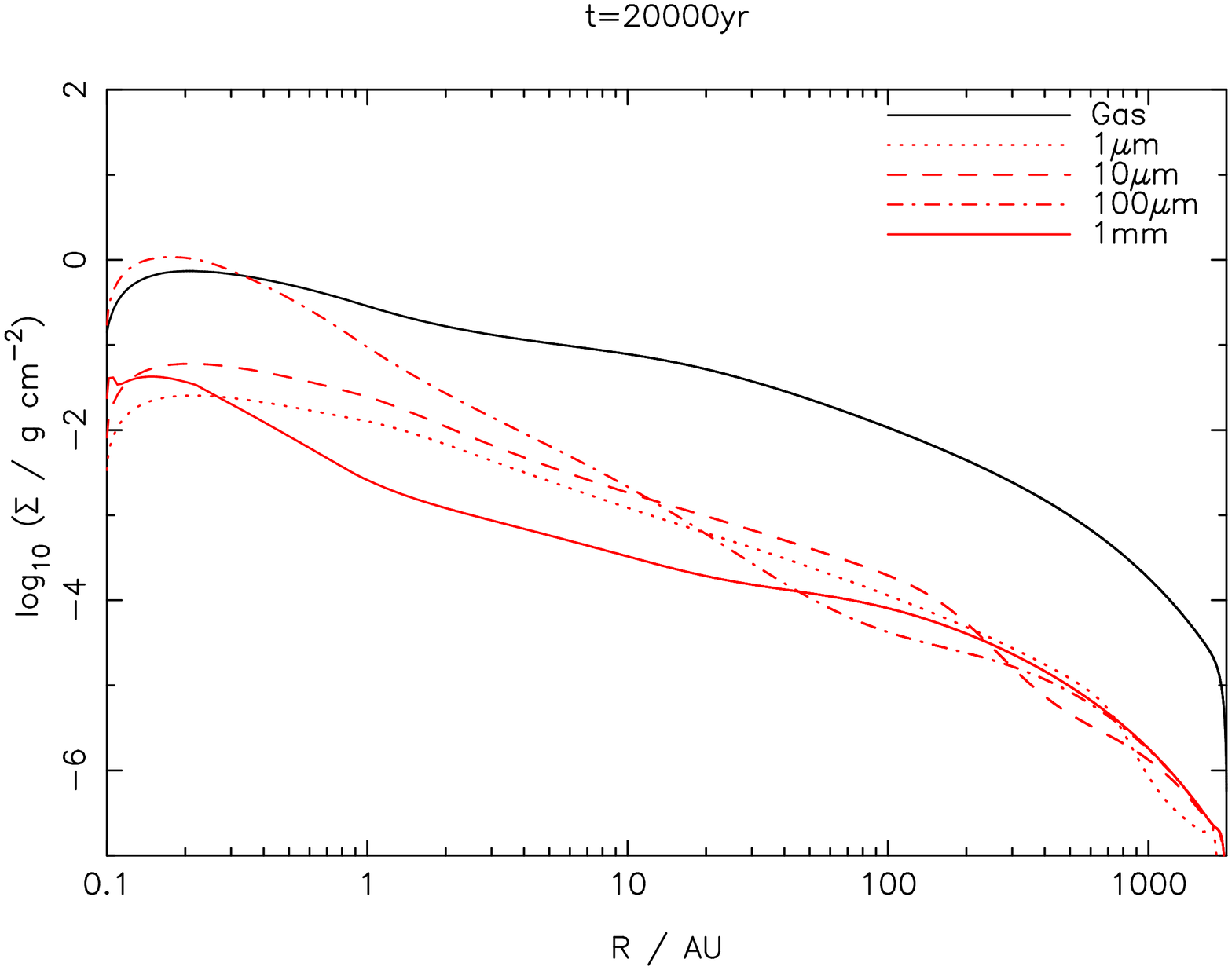}
         \includegraphics[angle=270]{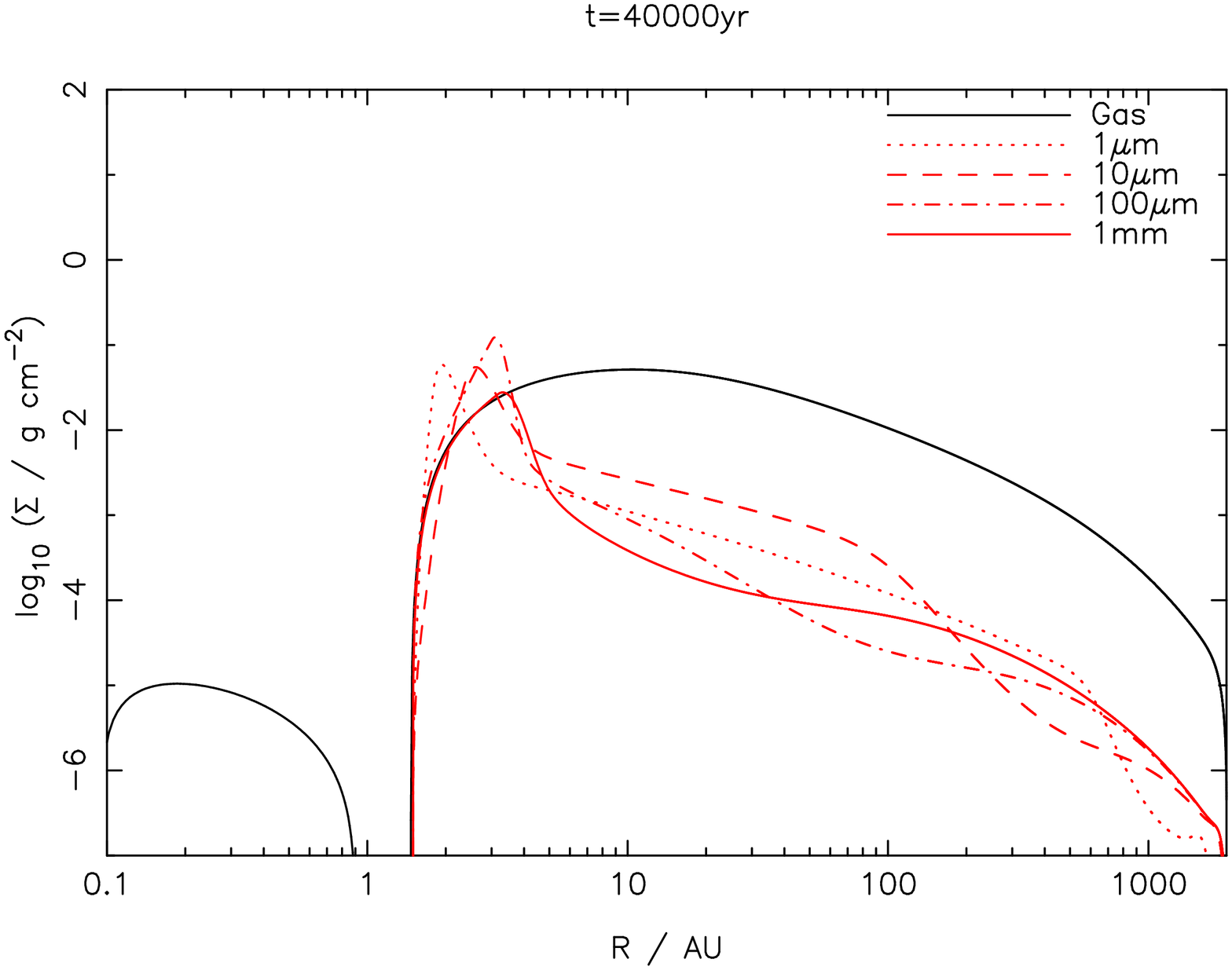}
        }
        \resizebox{\hsize}{!}{
        \includegraphics[angle=270]{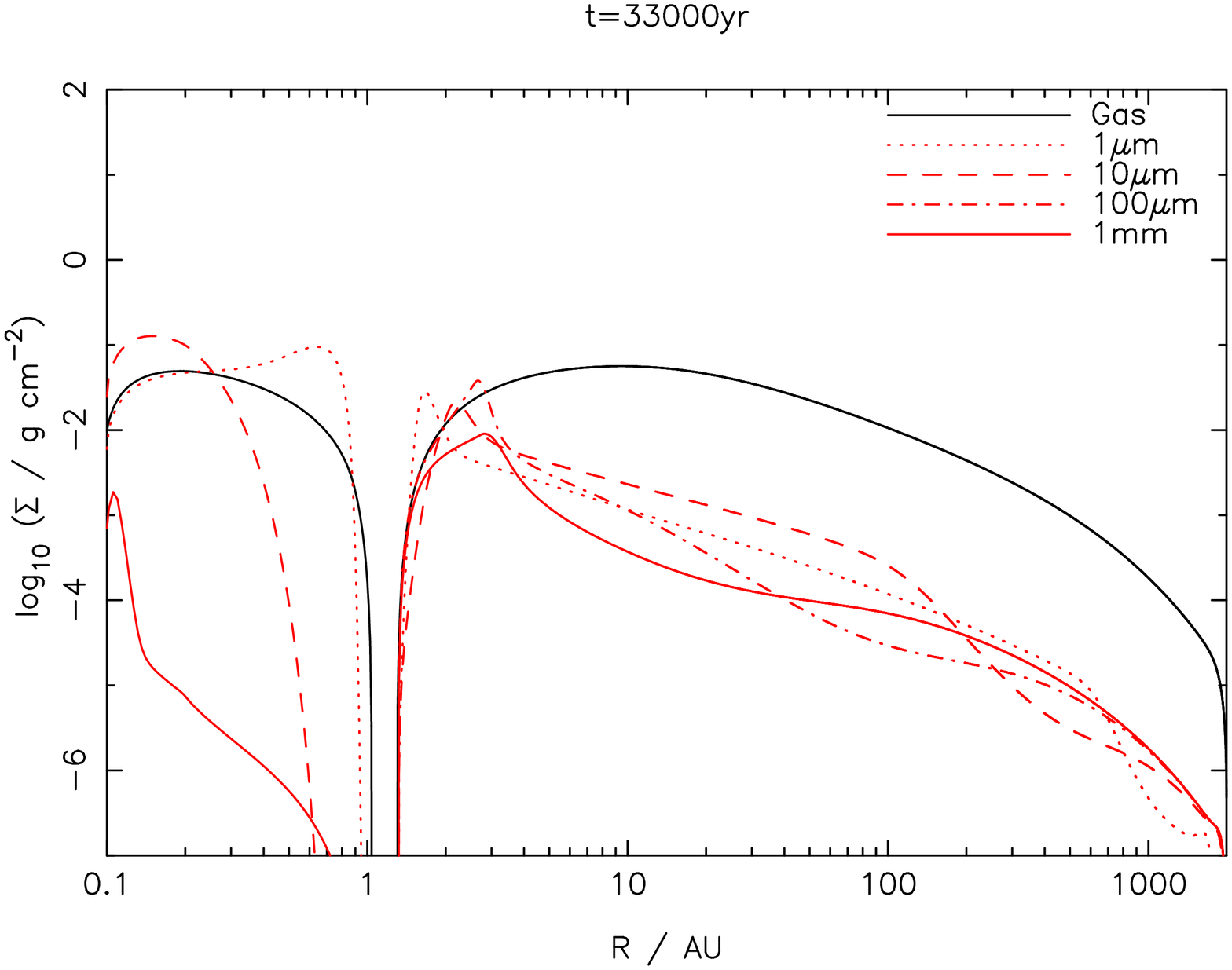}
        \includegraphics[angle=270]{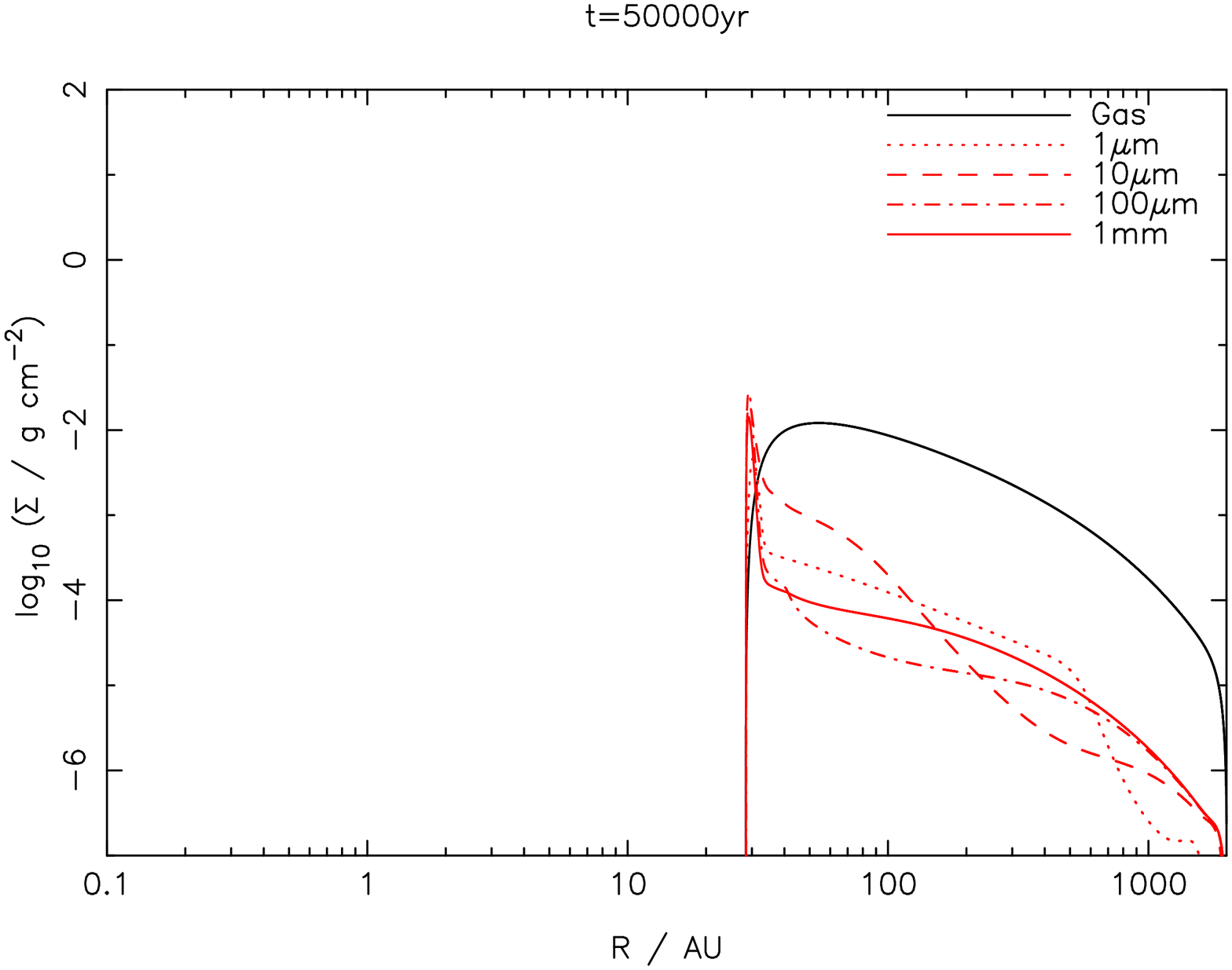}
        }
        \caption{Snapshots of the evolution of the gas and dust discs for the high-viscosity ($\alpha=0.01$) models: snapshots are plotted at $t=0$, 2.0, 3.3, 3.6, 4.0 \& $5.0\times10^4$yr.  The gas surface density is plotted in black, with the dust surface density in red: different line styles correspond to different grain sizes.  Note that these figures show the results of 4 independent models (with different grain size $s$), and are plotted together merely for comparison.}
        \label{fig:dust_high}
\end{figure*}

\begin{figure*}
\centering
        \resizebox{\hsize}{!}{
        \includegraphics[angle=270]{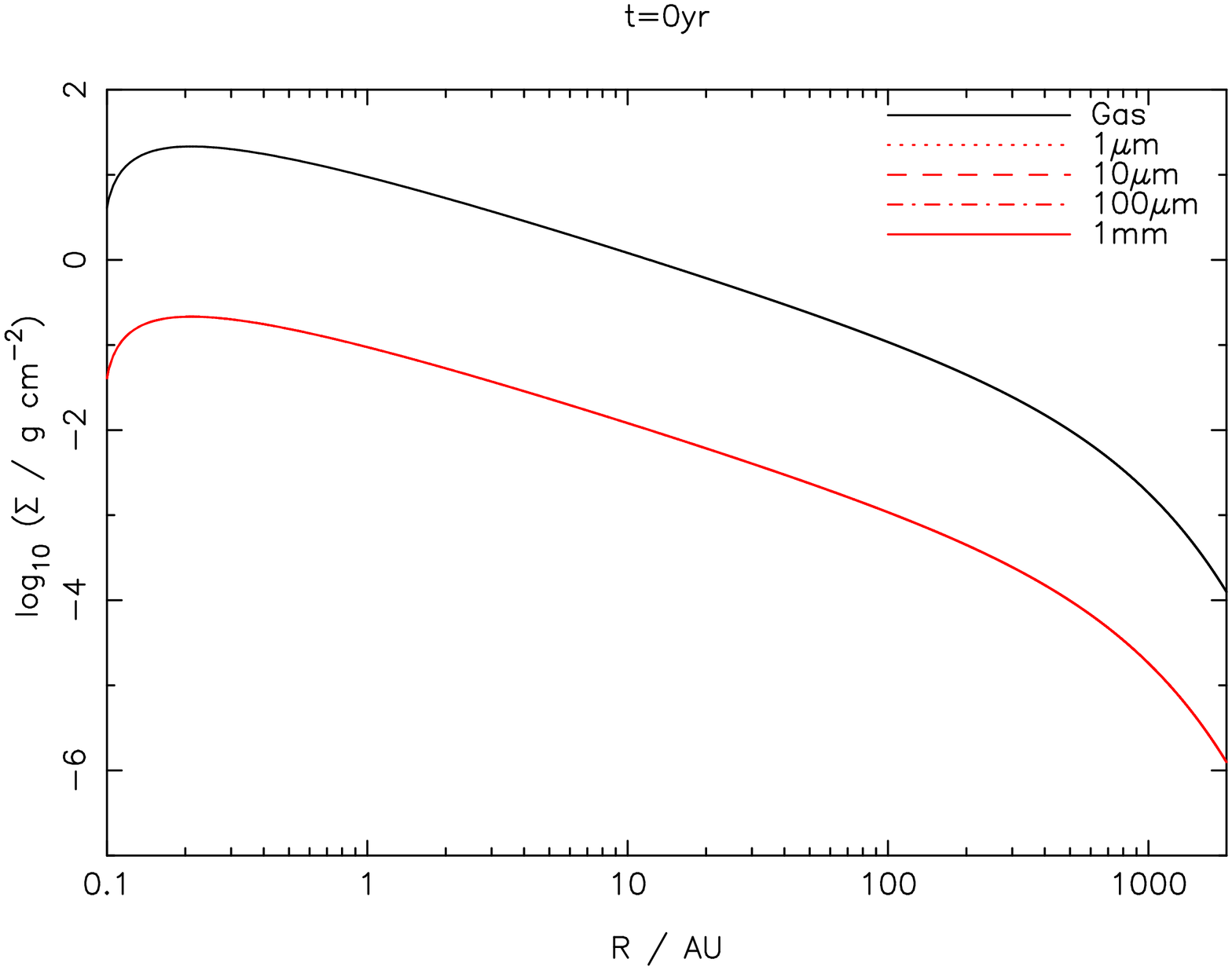}
        \includegraphics[angle=270]{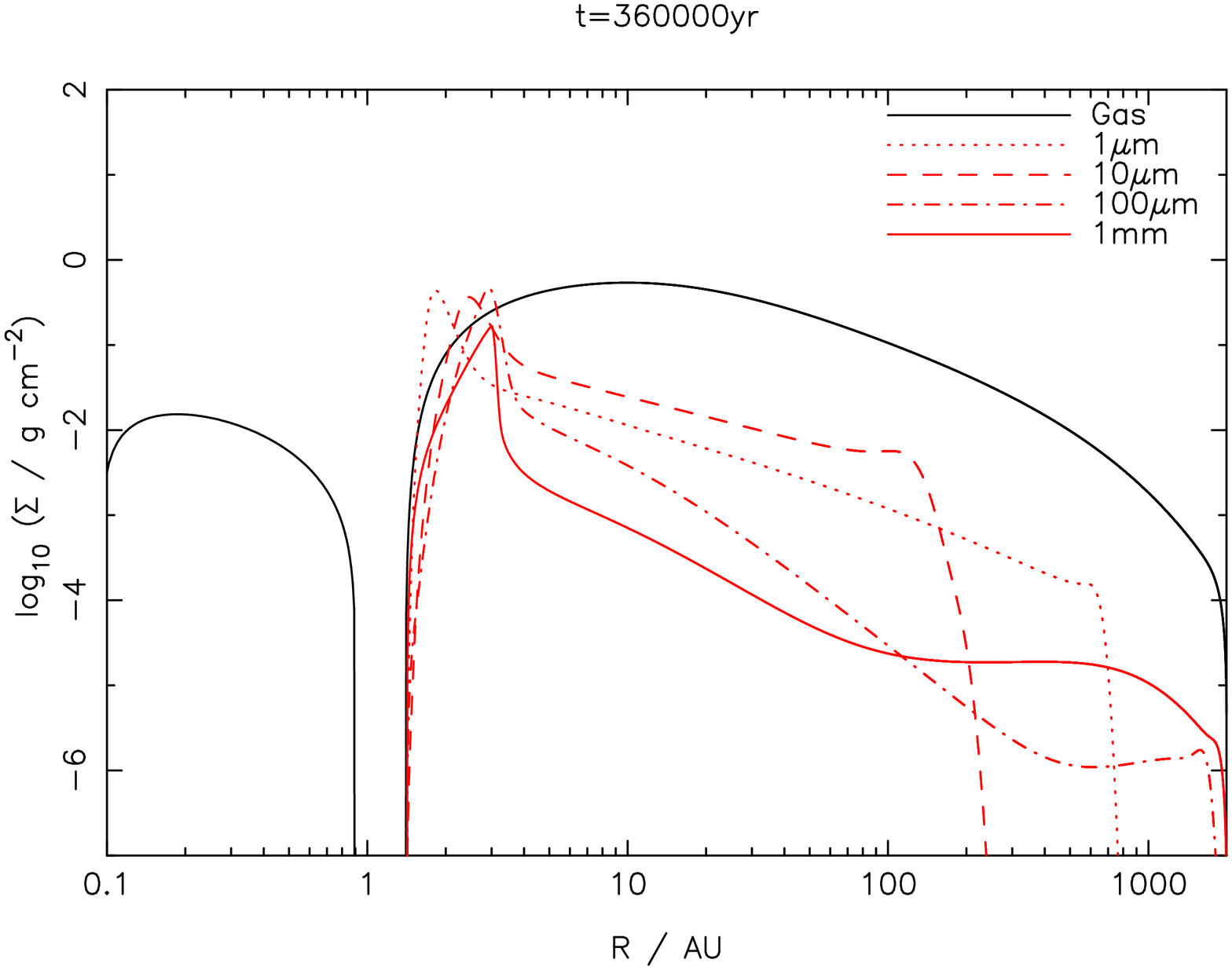}
        }
        \resizebox{\hsize}{!}{
        \includegraphics[angle=270]{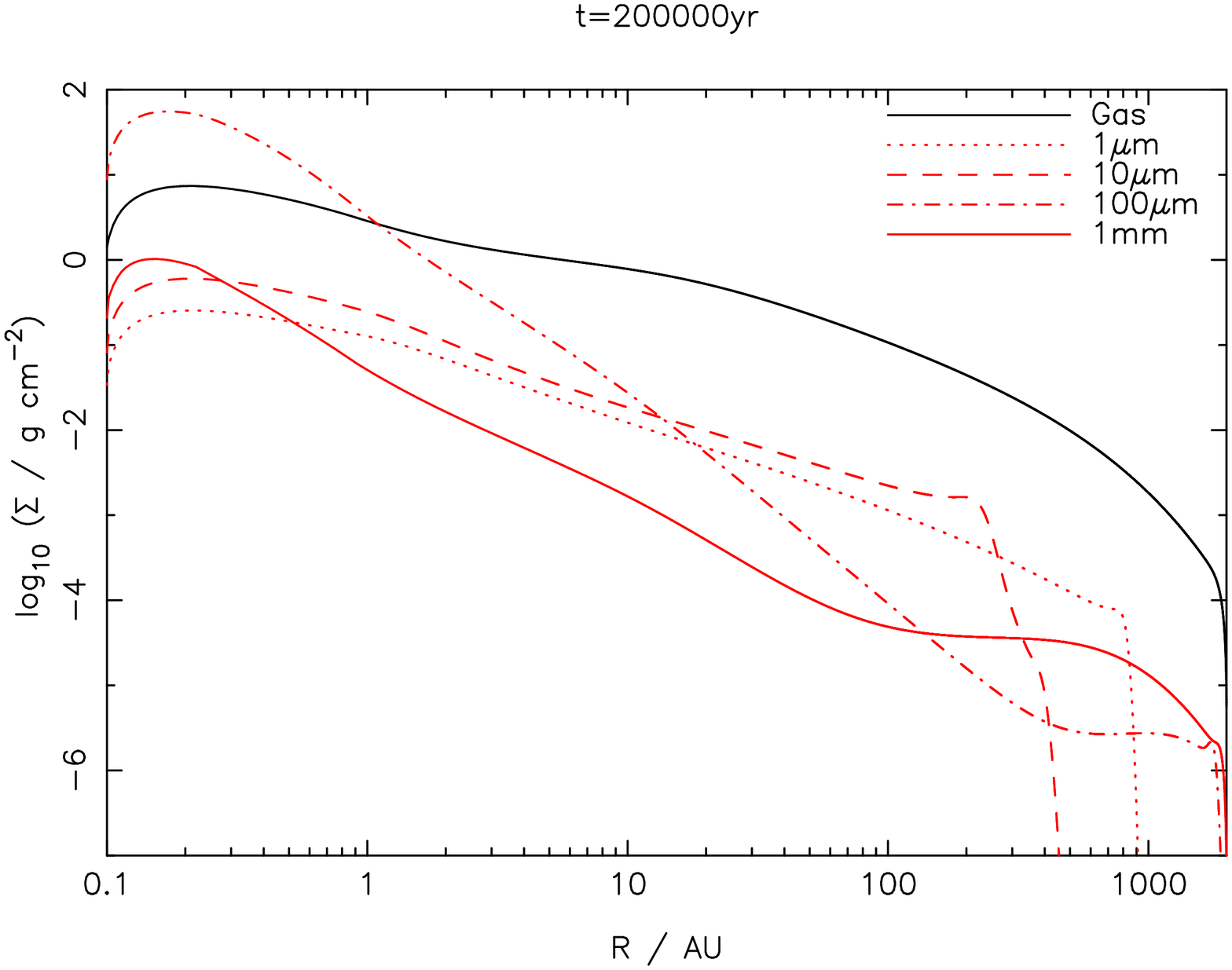}
        \includegraphics[angle=270]{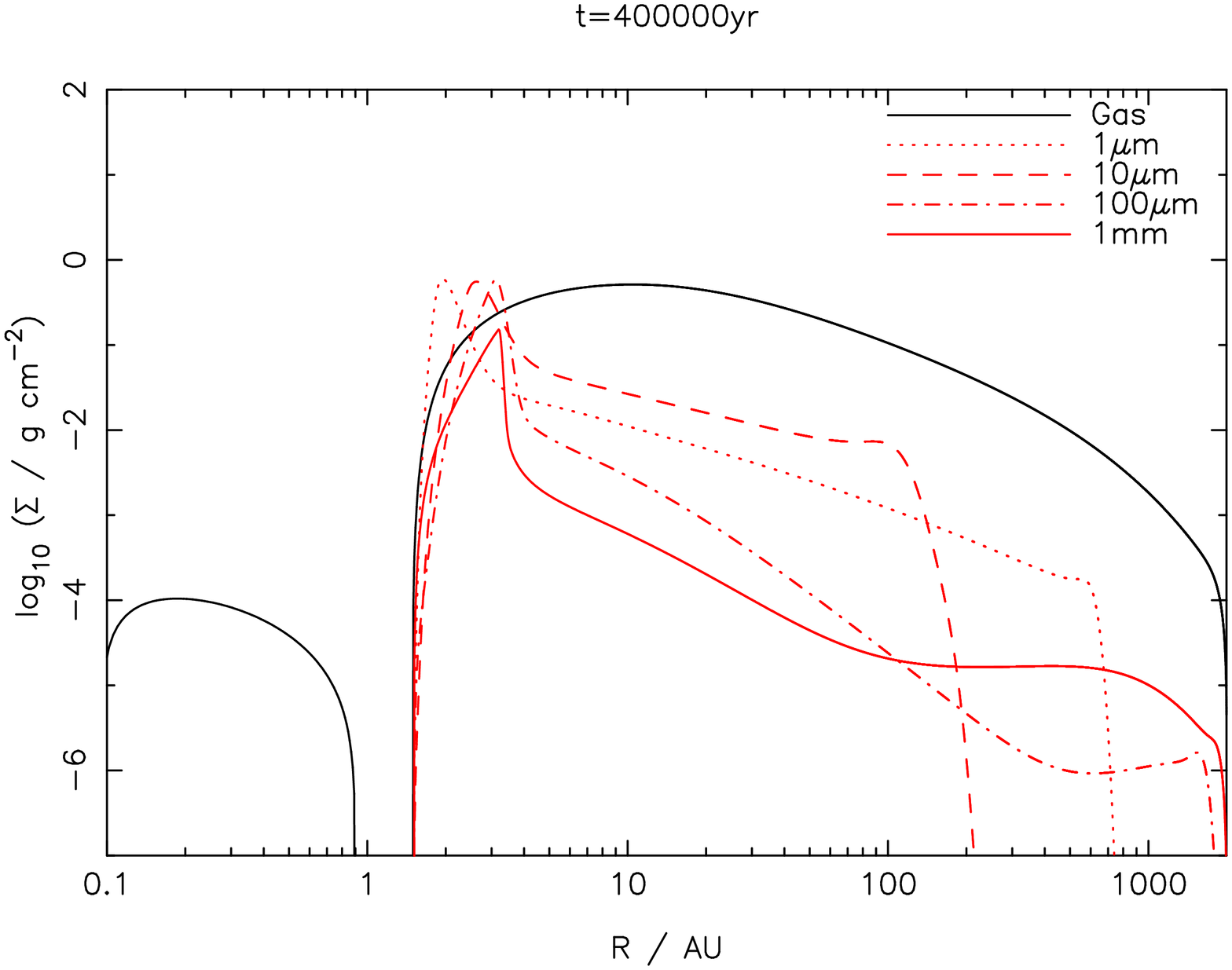}
        }
        \resizebox{\hsize}{!}{
        \includegraphics[angle=270]{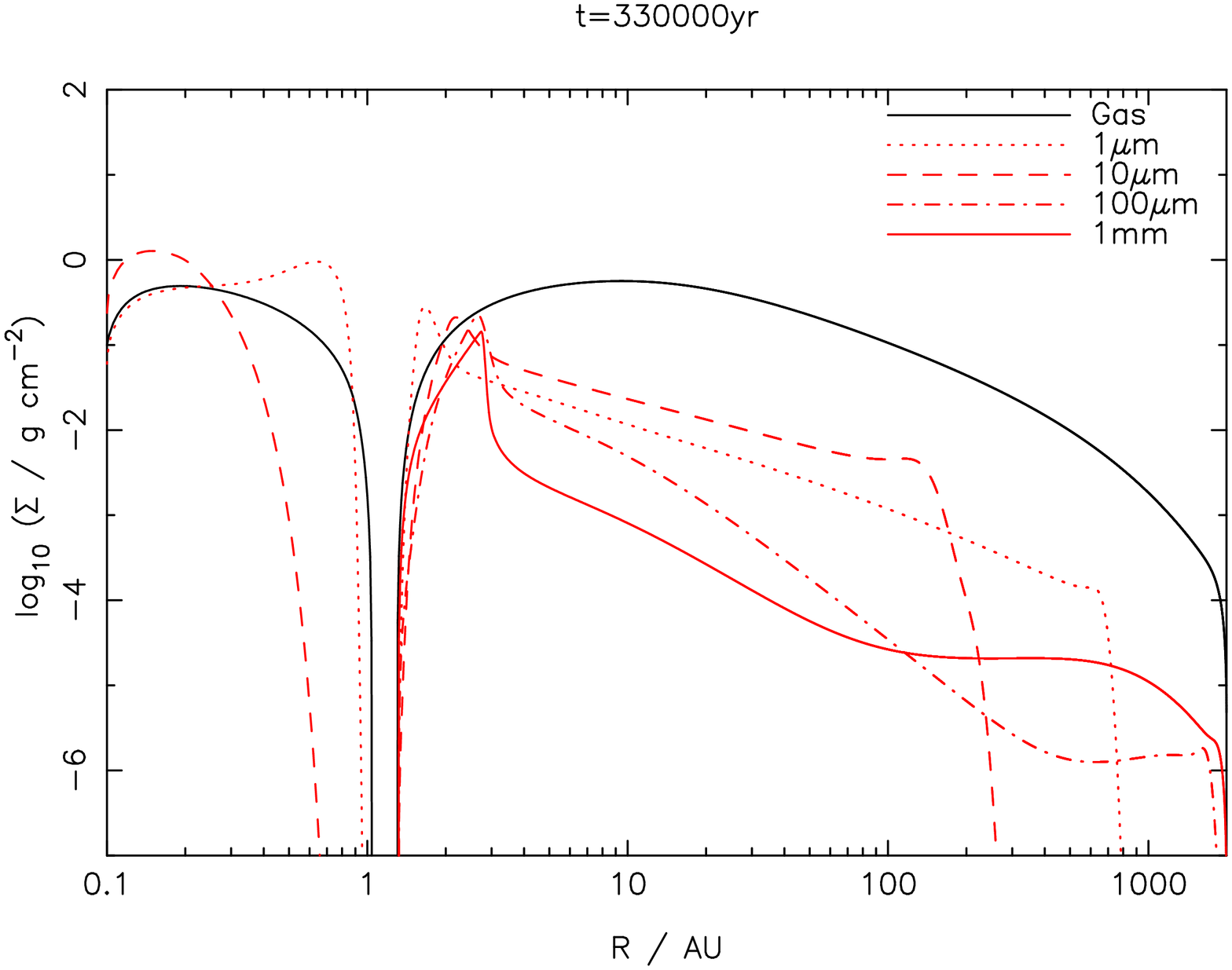}
        \includegraphics[angle=270]{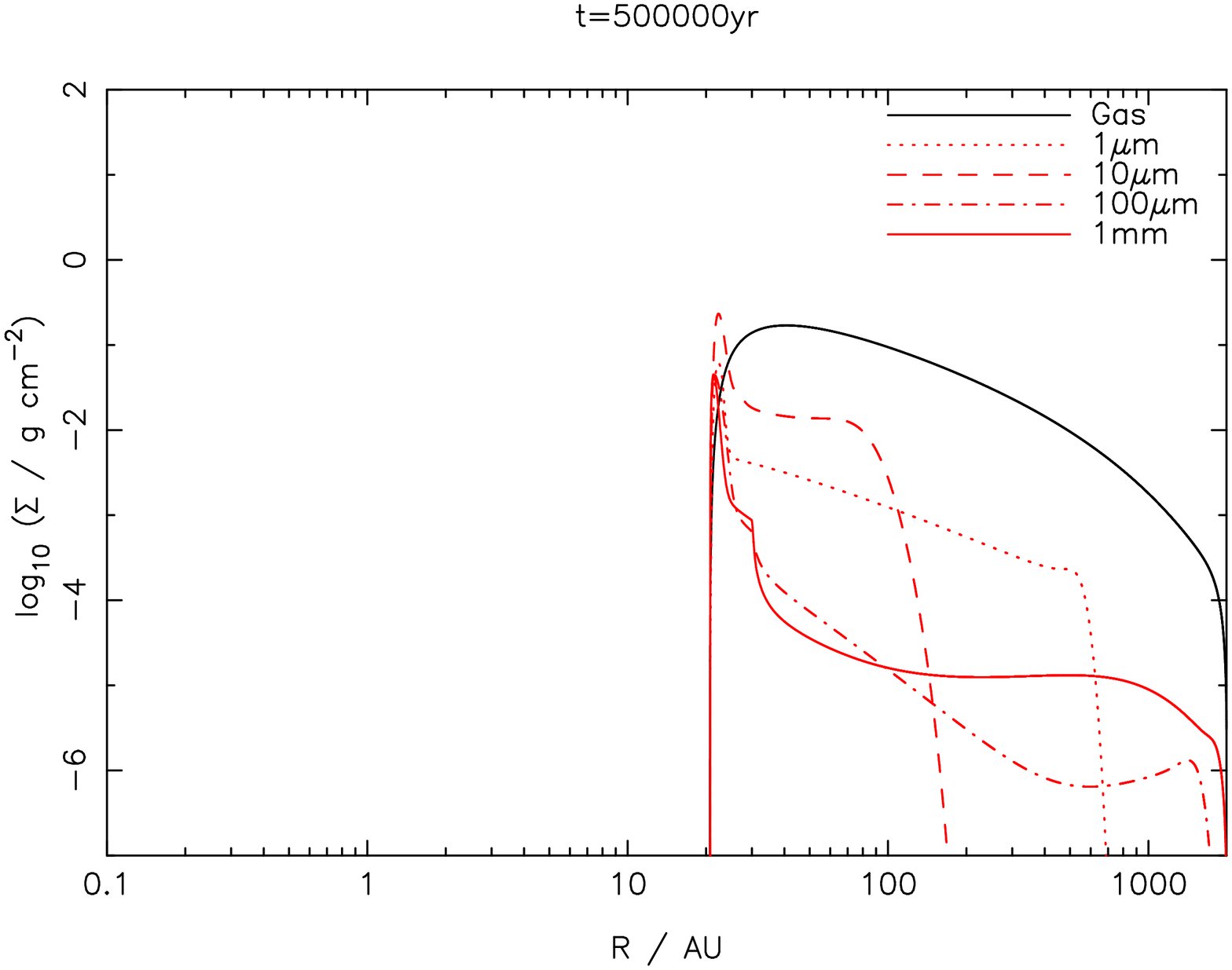}
        }
        \caption{As Fig.\ref{fig:dust_high}, but for the low-viscosity ($\alpha=0.001$) case.  Here the snapshots are plotted at $t=0$, 2.0, 3.3, 3.6, 4.0 \& $5.0\times10^5$yr.}
        \label{fig:dust_low}
\end{figure*}

Fig.\ref{fig:timescales} shows how the orbital decay time-scale (at $R=10$AU) due to advection, $t=-R/u_{\mathrm d}$, varies as a function of grain size $s$, in the initial disc.  Also shown is the (constant) orbital decay time-scale for the gas disc (essentially the viscous scaling time $t_{\nu}$).  As we have adopted $D=\nu$, the diffusive time-scale of the dust disc is simply given by 
\begin{equation}
t_{\mathrm {dif}} \simeq \frac{t_{\nu}}{\frac{d \log (\Sigma_{\mathrm d}/\Sigma_{\mathrm g})}{d \log R}} \, .
\end{equation}
We see, therefore, that the dust diffuses on a time-scale that is typically a factor of a few shorter than the viscous time-scale.  However, if the concentration gradient becomes large, then the dust diffuses much faster.  In our models the concentration gradient term is always greater than unity, but rarely becomes larger than $\sim5$ except close to the gap in the disc.  We therefore see from Fig.\ref{fig:timescales} that (prior to the opening of the gap) the evolution of small grains, with $s \lesssim 10$$\mu$m, is dominated by the diffusion, while the evolution of the larger grains is dominated by advection.  Thus we expect the concentration of small grains to remain approximately uniform up to the point where the wind opens the gap in the disc.  Larger grains, by contrast, migrate inwards more rapidly than the gas disc evolves, so in this case we expect significant evolution of the dust inside 10--20AU before the gap is opened.

Fig.\ref{fig:timescales} also highlights a key difference between dust evolution in the high- and low-viscosity models.  In the high-viscosity model the gap in the gas disc opens after $t=3.3\times10^4$yr, but this takes 10 times longer in the low-viscosity model.  The difference in the decay times for large ($\sim1$mm) grains, however, differs only by a smaller factor (around 1--5).  Consequently, we expect to see a more pronounced inward migration of such grains in the low-viscosity model.

Fig.\ref{fig:dust_high} shows the evolution of the four high-viscosity models\footnote{Note that we assume an initial dust-to-gas ratio of 0.01 in each model (i.e.~for each grain size), rather than a total dust-to-gas ratio of 0.01.  The absolute values of $\Sigma_{\mathrm d}$ correspond to the surface densities we would see if all the dust mass was contained in grains of a particular size: we plot the different models together in order to illustrate the differential migration of different grain sizes.}.  We see some inward migration of the larger grains at early times (before the gap opens), but we do not see a significant depletion of grains beyond $\sim 10$AU during this phase.  Once the gap opens, however, we see a very different behaviour.  The gas surface density increases sharply outside the gap (i.e.~at the inner edge of the outer disc), resulting in a strong inward pressure gradient.  The gas close to the gap edge is therefore super-Keplerian, and we see rapid {\it outward} migration of the grains.  Consequently, once the gap has opened the inner dust disc is cut off from resupply, and grains of all sizes drain rapidly onto the central star.  As the direct field photoevaporates the outer gas disc, the inner edge of the disc moves steadily outwards.  During this phase the steep pressure gradient at the inner edge continues to drive rapid outward migration of the grains, with the consequence that grains of all sizes are ``swept up'' by the moving edge of the gas disc.  As this process continues, more and more dust is accumulated into the region close to the edge of the gas disc, and eventually the dust-to-gas ration becomes large.  At this point our method, treating the dust as a trace contaminant, is no longer valid, so we stop our calculation at this point.  The ``final'' configuration is a roughly constant dust-to-gas ration at large radii for all grain sizes, with a concentrated dust ring at the inner disc edge.

The evolution of the low-viscosity models (as seen in Fig.\ref{fig:dust_low}), however, is rather different.  In this case the larger, advectively-dominated grains (100$\mu$m and 1mm) undergo much more significant radial migration before the gap opens.  Consequently, at the point where the gap opens, the outer disc is significantly depleted in large grains (out to radii $>100$AU), by an order of magnitude or more compared to the high-viscosity case.  Once the gap opens we again see rapid draining of dust from the inner disc, and again the inner edge of the gas disc sweeps up the dust as it moves outwards.  However the ``final'' state of the evolution shows a marked deficit of large grains, by comparison to the high-viscosity model.  We also note that the smaller $\mu$m-sized grains undergo significant inward migration from the outermost radii ($>500$AU).  This is due to the steepening pressure gradient beyond the ``disc radius''  $R_{\mathrm d}=500$AU, however, and is not physically significant (as it depends only on the choice of $R_{\mathrm d}$).

\section{Discussion}\label{sec:dis}
\subsection{Observational appearance of clearing discs}\label{sec:photo_holes}
A significant result from these models is that we are now able to make detailed predictions regarding the observable properties of so-called ``transitional discs'', which are ``caught in the act'' of clearing their discs.  Our models show that the dust-to-gas ratios for small grains are approximately constant (away from the edge of the gas disc), so our predictions regarding near- and mid-infrared emission agree well with previous gas-only models \citep[e.g.][]{acp06b}.  The gap is opened at $\simeq1$AU, and once the inner disc is drained the inner disc edge moves outward in radius.  For the viscosity law adopted here ($\nu \propto R$) the time required to double the hole radius scales as $R^{1/2}$, but a steeper viscosity law will result in a flatter distribution of hole sizes \citep[see Section 2 of][]{acp06b}.  Given the short duration of the clearing phase relative to the disc lifetime and the corresponding paucity of transition discs, it is unlikely that this weak bias towards larger hole sizes will be observationally significant.  Consequently, we predict that the distribution of inner hole sizes should be approximately uniform, from a minimum of $\simeq 1$AU outwards (for a 1M$_{\odot}$ star).  

As most of the disc mass resides at large radius, the disc mass remains approximately constant during clearing \citep[as seen in][]{acp06b}.  The exact value depends on the model parameters (viscosity, stellar ionizing flux and outer disc radius), but disc masses of a few Jupiter masses are typical.  We note, however, that disc masses are typically measured from dust emission at millimetre wavelengths, so these observations are subject to the uncertainties discussed in Section \ref{sec:mm}.

The constraints of the photoevaporation model require that the accretion rate on to the star be essentially zero during the inner hole phase, as once the inner disc is drained no accretion can persist.  During the inner disc draining, the accretion rate falls from $\simeq 10^{-10}$M$_{\odot}$yr$^{-1}$ to $<10^{-15}$M$_{\odot}$yr$^{-1}$ on a very short, viscous time-scale.  Consequently we predict that few, if any, objects should have measurable accretion rates below the photoevaporative wind rate of $\simeq 10^{-10}$M$_{\odot}$yr$^{-1}$, and note that the observational detection of this accretion rate threshold would provide strong support for the photoevaporation scenario.

Moreover, once the gap appears, the model predicts (and indeed requires) that the inner hole region, inside $\simeq1$AU, be essentially devoid of gas.  As noted in Section \ref{sec:gas_res}, in order for the inner hole to be optically thin to ionizing photons the total (neutral) gas mass inside 1AU must be $\lesssim 10^{-9}$M$_{\odot}$.  Recent observations with the {\it Spitzer Space Telescope} have placed strong upper limits on the gas mass in systems with cleared or clearing discs.  \citet{holl05} looked at the nearby star HD105, and placed upper limits of $\lesssim 0.01$M$_{\mathrm {Jup}}$ on the gas mass inside 1AU.  More recently, \citet{pasc06} observed 12 nearby stars with optically thin dust discs and concluded that there was $<0.0008$M$_{\mathrm {Jup}}$ ($<8\times10^{-7}$M$_{\odot}$) of gas in the region from 0.3--1AU.  These upper limits are rather close to the threshold value above, suggesting that the inner disc does indeed become optically thin to ionizing photons at late times, and adding further support to the photoevaporation model for disc clearing.

It is less clear, however, how the appearance of inner holes should scale with stellar mass.  In recent years, observations of disc properties, such as accretion rates and disc masses, have expanded to cover a much wider range in stellar mass than was previously possible \citep*[e.g.][]{muz05,scholz06}, and a number of trends with stellar mass are now known.  It has been suggested that the fraction of ``transition discs'' (identified from near- and mid-infrared observations of SEDs) increases with decreasing stellar mass: both \citet{mccabe06} and \citet{lada06} found that ``SED holes'' were more prevalent among later type stars.  The degree to which these results are influenced by selection biases remains unclear, as these data are only sensitive to ``holes'' at $\lesssim 20$$\mu$m, but we note in passing that such an effect occurs naturally in our models if the viscous time-scale increases with decreasing stellar mass \citep[as suggested by][]{aa06}.

\subsection{Dust at the inner disc edge}\label{sec:edge}
An interesting result of our model is the behaviour of the dust close to the inner edge of the gas disc.  As outlined above, once the gap opens the dust in the inner disc is rapidly drained onto the star.  Outside the gap, however, the inward pressure gradient results in outward migration of the dust grains, and inward accretion of dust is halted at, or close to, the peak in the surface density profile.  Consequently the inner disc cannot be resupplied with dust, resulting in a very ``clean'' inner hole.

Once the inner gas disc is drained, the disc edge begins to move outward.  The steep pressure gradient at the edge causes the grains to be ``swept up'' as the gas edge moves outwards, resulting in a rather concentrated, dust-rich, ring immediately outside the inner edge of the gas disc.  As seen from Figs.\ref{fig:dust_high} \& \ref{fig:dust_low}, the behaviour of this dust ring is largely independent of grain size over the range we consider.  The exact form of this ring in our models depends rather strongly on the wind profile at the inner disc edge (defined by the smoothing term in Equation \ref{eq:smooth}), but such a ring occurs for any choice of parameters.  Moreover, similar structures, caused by the same physical process (a reversal in the direction of the pressure gradient close to the disc edge), have been found by other authors \citep[e.g.][]{pm04,pm06}, so although we are unable to make detailed predictions we are satisfied that the formation of the dust-rich ring is a real physical process. 

As the disc edge sweeps outwards the dust-to-gas ratio close to the edge rises, and at late times can exceed unity.  At this point our two-fluid treatment of the disc breaks down (so we stop the models at this point), and the evolution of the system becomes rather uncertain.  Beyond this point the dust dynamics will be largely independent of the gas disc, and it may well be that such a structure is unstable \citep[e.g.][]{gw73,gl04}.  Alternatively, the enhanced dust concentration may result in enhanced grain collision rates, and may facilitate rapid grain growth through coagulation.  Again, the details are somewhat model-dependent but, given that gas is ``lost'' to the wind while dust is merely redistributed radially, it is straightforward to estimate at what radius the dust-to-gas ratio of the ring will reach these large values: this typically occurs at $\sim 50$--100AU.  Other studies have similarly found that local pressure maxima can result in significantly enhanced dust concentrations \citep{kl01,kl05,rice06b,bw06}, and speculated as to the possible consequences in terms of planetesimal and debris disc formation.  Detailed study of these consequences are not possible here, but we note that the formation of the dust-rich ring during the gas clearing phase can in principle lead to rapid growth of planetesimals, and may result in the formation of a debris disc at a radius of $\sim$50AU.

Were we to extend our analysis to larger grain sizes, however, it seems likely that grains larger than some critical size would be so weakly coupled to the gas that they would not be swept up in this manner.  Inspection of Fig.\ref{fig:timescales} suggests that this critical size is likely in the range 10--100cm, but this is difficult to quantify without making detailed calculations.  ``Filtration'' of the dust in this manner is analogous to the process considered by \citet{rice06}.  They consider a different situation, namely the flow of gas and dust past a planet-induced gap at a fixed radius, but the underlying physical processes are the same.  The key difference between the the situation considered by \citet{rice06} and that considered here is that accretion of gas persists across a planet-induced gap, whereas a photoevaporative wind prevents the inward accretion of gas at the disc edge.  Consequently, in the case of a planet-induced gap small grains (which are strongly coupled to the gas) are carried across the gap by the gas, while in our models these grains pile up close to the disc edge.  Consequently, it seems that dust properties can be used to discriminate between different mechanisms for creating ``inner holes'': planet-induced gaps tend to populate the inner hole with small, micron-sized grains while filtering out larger particles, while photoevaporation removes all but very large grains (cm-size or larger) from the inner disc.

\subsection{Grain growth and replenishment}\label{sec:mm}
A significant simplification in our model is the fact that we neglect grain collisions.  Collisions between grains provide a means of altering the size distribution of grains, both through coagulation and destructive collisions \citep[e.g.][]{dominik06}.  Models have shown that the time-scales for grain growth and destruction can be significantly shorter than disc lifetimes, and have also shown that grains of all sizes must be subject to continuous resupply in order to sustain the observed grain populations over the lifetime of the gas disc \citep[e.g.][]{tcl05,dd05}.  Recent observations have shown evidence for dust settling and moderate grain evolution over disc lifetimes \citep{furlan06}, but wholesale changes in the grain population do not occur.

Our model considers only the short, clearing phase of the evolution, but even over these short time-scales we see significant inward migration of 100$\mu$m- and mm-size grains from large radii ($>50$AU).  Observations require that this grain population be sustained even during the clearing phase, and this enables us to make crude estimates of the time-scale of grain growth in the outer disc.  Our high-viscosity model shows only moderate depletion of mm grains in the outer disc, but the depletion is much more significant in the  low-viscosity model.  This suggests that the time-scale for the replenishment of the mm grain population at radii of $\sim50$--100AU is $\lesssim10^5$yr.

\subsubsection{Simple replenishment model}
\begin{figure}
\centering
        \resizebox{\hsize}{!}{
        \includegraphics[angle=270]{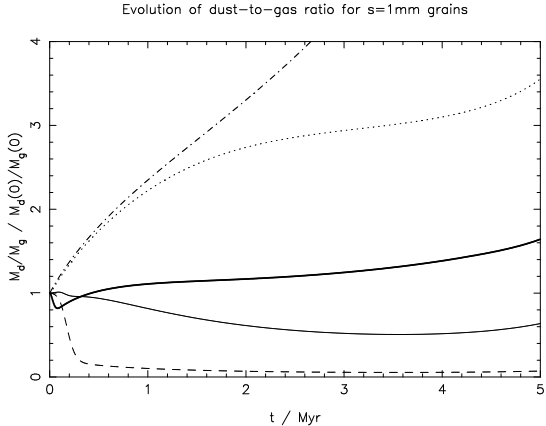}
        }
        \caption{Evolution of dust-to-gas ratio of $s=1$mm grains, normalised to the initial value, for various values of $t_{\mathrm {rep}}$.  The dot-dashed, dotted, solid and dashed lines are the results for $t_{\mathrm {rep}}=10^3$yr, $10^4$yr, $10^5$yr and $10^6$yr respectively.  The heavy solid line shows the results for $t_{\mathrm {rep}} = 200\pi/\Omega$ (i.e.~100 local orbital periods), which gives $t_{\mathrm {rep}} =10^5$yr at $R=100$AU.  We see that replenishment time-scales in the range $10^4$--$10^5$yr are required to sustain an approximately constant dust-to-gas ratio over Myr time-scales.
        }
        \label{fig:replen}
\end{figure}
In order to quantify this analysis, we have considered a model which includes a simple replenishment term.  We assume that the dust mass for a given grain size can be increased by growth processes, and add a source term to Equation \ref{eq:dust}.  The source term (which appears on the right-hand side of Equation \ref{eq:dust}) takes the form
\begin{equation}
\dot{\Sigma_{\mathrm d}}(R,t) = -\frac{\Sigma_{\mathrm d}(R,t)-\Sigma_{\mathrm d}(R,0)}{t_{\mathrm {rep}}} \, ,
\end{equation}
where $t_{\mathrm {rep}}$ is the time-scale for replenishment.  We consider a model of the entire disc lifetime, rather than just the clearing phase, with parameters $\alpha=0.01$, $R_{\mathrm d}=50$AU, and $\dot{M}_0 = 5\times10^{-8}$M$_{\odot}$yr$^{-1}$.  With these parameters the accretion rate falls to $10^{-10}$M$_{\odot}$yr$^{-1}$ after 4.8Myr, and the gap in the gas disc begins to open at 5.2Myr.

Fig.\ref{fig:replen} shows the evolution of the dust-to-gas ratio for models with $s=1$mm and various values of $t_{\mathrm {rep}}$.  We see that a replenishment time-scale of $10^5$yr results in an almost constant dust-to-gas ratio over the Myr lifetime of the disc.  Longer replenishment time-scales result in significant depletion of dust, while significantly shorter replenishment time-scales result in a dust-to-gas ratio which increases with time.  This provides an additional lower limit to the replenishment time-scale for mm grains in the outer disc ($\sim50$--100AU), and tells us that these grains must be replenished on a time-scale of $10^4$--$10^5$yr, in broad agreement with models of collisional grain growth \citep[e.g.][]{dd05}.  Moreover, this provides an {\it a posteriori} justification of our choice of initial conditions, and suggests that replenishment processes can indeed sustain a quasi-equilibrium grain population over the lifetime of the T Tauri phase.

\subsection{Limitations}
In addition to the issues discussed above, there a number of limitations to our analysis which bear consideration.  One potentially important physical mechanism which is neglected here is the effect of radiation on the dust grains, which in principle could result in either Poynting-Robertson (PR) drag or radiation-pressure blowout.  However, both of these effects are only usually significant in the absence of a gas disc, as the gas acts both to stabilise the dust against radiation forces, and to shield the grains from the radiation field. In our models the dust is always co-located with the gas disc, so PR drag is negligible.  Close to the inner edge of the disc radiation pressure may become significant, but the likely outcome of this process is the either radiative blow-out of the grains or the formation of a dust-rich ring close to the edge of the gas disc \citep{kl01}.  Our models already form such structures, due to the pressure gradients in the gas disc, so it is unlikely that the inclusion of radiation pressure will modify our results significantly.

In addition to these physical limitations, there are also numerical limitations to consider.  Our treatment of the dust close to the disc edge is rather crude (see Section \ref{sec:numerics}), but it is sufficiently accurate for situations we consider.  More significant, however, may be the form of the inner boundary condition.  We adopt zero-torque boundary conditions throughout, but note that these can result in outward migration of the grains near to the boundary (see Section \ref{sec:numerics}).  In order to minimise the influence of the boundary condition, we neglect the pressure terms in the region close to the inner boundary.  The behaviour of discs close to the inner boundary is not well understood, but some physically-motivated boundary conditions, such as magnetospheric truncation \citep*[e.g.][]{gl78,hhc94}, do result in conditions not dis-similar to those adopted here.  Consequently, in such cases dust grains may indeed ``pile up'' in the region close to the inner boundary.  We note, however, that in CTTs the inner edge of the gas disc usually lies inside the dust sublimation radius \citep{dalessio05b,eisner05}, so this effect is probably not significant in TT discs.

An interesting extension of this work would be to consider a range of grain sizes simultaneously.  Such a model would enable a more detailed analysis of the effects of grain growth and destruction than are possible here.  It would also allow us to consider the effects of factors such as the grain size distribution, and to study of the evolution over longer time-scales than those considered here.  This is obviously beyond the scope of this investigation, but provides an interesting avenue for future work.


\section{Discriminating between models}\label{sec:holes}
In light of these results, we now seek an observational diagnostic that will allow us to discriminate between the various different models for ``inner hole'' discs.  We propose that the simplest such diagnostic lies in the relative values of the disc mass and stellar accretion rate.  In order to compare to the results of our photoevaporation models, we now consider a simple model for a planet embedded in a disc.

Several studies have shown that a relatively massive planet can produce an observable ``hole'' in a protoplanetary disc \citep[e.g.][]{rice03,quillen04,rice06,varniere06}.  In these models the tidal interaction between the planet and the disc opens a gap in the disc, and the tidal barrier limits accretion across this gap.  In general this only happens if the planet mass is $\gtrsim 1$M$_{\mathrm {Jup}}$ \citep*[e.g.][]{al96,tml96,lsa99}, as less massive planets are not capable of opening such a gap unless the disc is unusually inviscid.  Moreover, the accretion rate past the planet is a strong function of the planet mass.

Our ``toy'' model is defined as follows.  We assume that the disc is in a steady state, and adopt the same viscosity law as used previously (see Section \ref{sec:visc}).  The total disc mass $M_{\mathrm d}$ is therefore given by
\begin{equation}\label{eq:disc_mass}
M_{\mathrm d} = \int_{R_{\mathrm {in}}}^{R_{\mathrm {out}}} 2 \pi R \Sigma(R) dR \, ,
\end{equation}
where $R_{\mathrm {out}}$ is the outer disc radius.  In a steady disc, the surface density can be expressed in terms of the steady disc accretion rate, $\dot{M}$, and the viscosity thus \citep{pringle81}
\begin{equation}
\Sigma(R) = \frac{\dot{M}}{3\pi\nu(R)} \, .
\end{equation}
We can integrate Equation \ref{eq:disc_mass} to find
\begin{equation}
\dot{M} = \frac{3 \alpha \Omega H^2}{2 R (R_{\mathrm {out}}-R_{\mathrm {in}})} M_{\mathrm d} \, ,
\end{equation}
and we note that $\Omega H^2/R$ is constant for our choice of the flaring index, $q$.  (A different choice of $q$, however, would not alter the results significantly.)  

We now seek to relate the steady disc accretion rate, $\dot{M}$, to the observed accretion rate on to the stellar surface, $\dot{M}_{\mathrm {acc}}$.  This depends on the flow of material across the gap induced by the planet, and is not yet fully understood.  The models of \citet{lda06} suggest that the accretion rate past the planet (into the inner disc) is typically 10--25\% of the accretion rate on to the planet.  The accretion rate on to the planet, $\dot{M}_{\mathrm p}$, in turn depends on the planet mass $M_{\mathrm p}$.  This interaction is still the subject of some debate, but for simplicity we adopt the approximate scaling derived by \citet{lsa99} 
\begin{equation}
\dot{M}_{\mathrm p} \simeq \left(\frac{M_{\mathrm p}}{M_{\mathrm {Jup}}}\right)^{-1} \dot{M} \, , \quad M_{\mathrm p} \ge M_{\mathrm {Jup}}
\end{equation}
and, following \citet{lda06}, set 
\begin{equation}
\dot{M}_{\mathrm {acc}} \simeq 0.25 \dot{M}_{\mathrm p} \, .
\end{equation}
Consequently, the observed accretion rate and the disc mass are related by
\begin{equation}
\dot{M}_{\mathrm {acc}} \simeq 0.25 \left(\frac{M_{\mathrm p}}{M_{\mathrm {Jup}}}\right)^{-1}  \frac{3 \alpha \Omega H^2}{2 R (R_{\mathrm {out}}-R_{\mathrm {in}})} M_{\mathrm d} \, ,
\end{equation}
which is a function only of the planet mass, viscosity and disc size.  For simplicity, we fix $R_{\mathrm {out}}=100$AU and consider a range of planet masses and viscosities.  We consider planet masses ranging from 1$M_{\mathrm {Jup}}$ (approximately the minimum mass required to open a gap in the disc) to 12$M_{\mathrm {Jup}}$ (approximately the deuterium-burning limit), and viscosity parameters ranging from $\alpha=0.01$--0.001.

\begin{figure}
\centering
        \resizebox{\hsize}{!}{
        \includegraphics[angle=270]{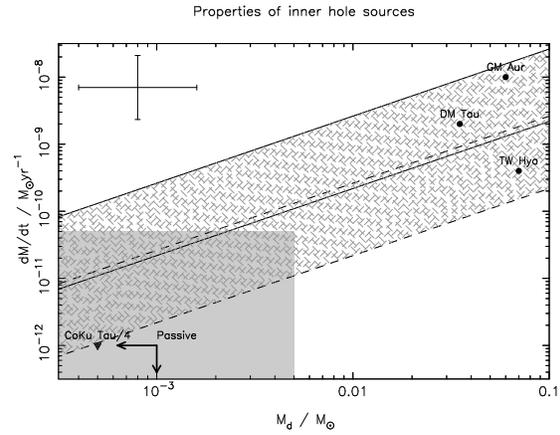}
        }
        \caption{Disc mass versus accretion rate for different models of inner hole sources.  The solid lines show the result of the ``planet'' model: solid lines show the results for $\alpha=0.01$; dashed lines denote $\alpha=0.001$.  In each case the upper and lower lines show the results for the minimum (1M$_{\mathrm {Jup}}$) and maximum (12M$_{\mathrm {Jup}}$) planet masses respectively.  The hatched region therefore represents the range of parameters allowed by the planet model.  The solid grey region shows the range allowed by the photoevaporation model but, as noted in the text, most such inner hole sources are not expected to have detectable accretion.  Also shown are data for several known inner hole sources, as described in the text: circles represent detections, while triangles and arrows denote upper limits.  The error bar in the upper left corner shows the typical systematic error in the observational derivation of these parameters.}
        \label{fig:holes}
\end{figure}

Fig.\ref{fig:holes} shows the range of parameters allowed by this model, as well as that of the photoevaporation model.  As discussed in Section \ref{sec:photo_holes}, objects appear as inner hole sources in the photoevaporation model only when the accretion rate falls below the wind rate, and unless the disc is unusually inviscid, or unusually large, this implies a maximum disc mass of $\simeq0.005$M$_{\odot}$.  In principle it is possible to observe a ``photoevaporated'' inner hole with a detectable accretion rate (as in the fourth snapshot of Fig.\ref{fig:dust_low}), but the window of opportunity for such a detection is very short and the majority of such objects have no significant accretion.  By contrast, planet-induced inner holes tend to show significant accretion rates, and can occur over a range of disc masses.  For a given disc mass a more massive planet results in a lower accretion rate, and a lower viscosity results in a similarly lower accretion rate.

Also shown in Fig.\ref{fig:holes} are the properties of several well-studied transition objects: GM Aur, DM Tau, TW Hya and CoKu Tau/4.  These data are taken from \citet[][TW Hya]{calvet02} and \citet[][GM Aur \& DM Tau]{calvet05}, with disc masses for sources in Taurus-Auriga taken from the survey of \citet{aw05}.  CoKu Tau/4 is a spectroscopic WTT with no evidence of on-going accretion so, following \citet{sic06}, we assign an upper limit to the accretion rate of $10^{-12}$M$_{\odot}$yr$^{-1}$.  In addition, we have included the locus of the ``passive discs'' identified in the survey of binary objects conducted by \citet{mccabe06}.  This survey identified six inner hole sources, four of which are in Taurus-Auriga.  \citet{aw05} derived disc masses of $\simeq0.0005$--0.001M$_{\odot}$ for these objects, but were not able to resolve the individual components of these binary systems.  Consequently we treat these measurements as upper limits to the disc masses, as they essentially measure the total disc masses of the binary pairs.  All of these objects are WTTs, so we again assign an upper limit of $10^{-12}$M$_{\odot}$yr$^{-1}$.

This toy model obviously has uncertainties of factors of a few, but despite this Fig.\ref{fig:holes} suggests that observations of disc masses and accretion rates potentially provide a powerful tool for discriminating between different models.  GM Aur, DM Tau and TW Hya fall in the region predicted by the planet model, while CoKu Tau/4 and the objects identified by \citet{mccabe06} are more readily explained by photoevaporation.  The reason that this simple method can discriminate so strongly between these models is easily understood.  We select objects based on the ``transitional'' appearance of their SEDs only, but then consider parameters that relate to the global disc structure.  An embedded planet affects the SED dramatically, but has only a modest (factor of $\sim$10--100) effect on the accretion rate and little effect on the total disc mass (as the most of the disc mass resides at large radii).  By contrast, photoevaporative clearing requires significant changes in the global disc properties, and therefore occupies a markedly different region of parameter space in Fig.\ref{fig:holes}.

We note, however, that our toy planet model does not consider the formation of the planet, but merely assumes its presence.  More detailed analysis of this problem can provide additional insight, and consideration should also be given to the evolution of the planet-disc system over $\sim$Myr time-scales.  Additional parameters, such as stellar ages and masses, can also be used to constrain models further and more detailed observations of disc properties can add valuable additional constraints \citep[e.g.][]{eisner06}.  However, our results suggest that even simple, easily observed parameters, such as disc masses and accretion rates, can provide a strong discriminant between different models of disc clearing.

\section{Summary}\label{sec:summary}
In this paper we have studied the dynamics of dust and gas during protoplanetary disc clearing.  We have constructed two-fluid numerical models, in which the gas is subject to viscous accretion and photoevaporation and the dust undergoes both radial diffusion and advection.  We find that small grains ($\lesssim 10$$\mu$m) grains remain well-coupled to the gas throughout, while larger grains undergo significant radial migration over the duration of the clearing phase.  This suggests that mm grains in the outer disc must be replenished on time-scales of $10^4$--$10^5$yr, presumably through collisional growth.  Once the photoevaporative wind opens a gap in the gas disc the dust is rapidly removed from the inner disc.  As the outer gas disc is cleared by the wind the grains are ``swept up'' by the steep pressure gradient at the inner edge of the gas disc, resulting in a dust-rich ring near to the inner disc edge.  This may result in increased rates of grain collision and growth, with the potential to form planetesimals and/or a debris disc.  We have described in detail the observational appearance of discs during the clearing phase in our models, and find that several observed discs agree well with the predictions of our model.  Lastly, by considering a simple model of a planet embedded in a disc we have outlined how observations of the masses and accretion rates of transitional discs can provide a means of discriminating between different models of disc clearing.

\section*{Acknowledgements}
We are grateful to Cathie Clarke for a number of useful discussions and a critical reading of the manuscript.  We also acknowledge interesting discussions with both Ken Rice and Carsten Dominik, and RDA thanks the IoA, Cambridge for hospitality during a visit.  We thank Ian McCarthy for providing numerical results from the hydrodynamic models of \citet{font04}, and are grateful for permission to reproduce their numerical fit in the Appendix.  Lastly, we thank an anonymous referee for several useful comments which improved the clarity of the paper.

This work was supported by NASA under grant NNG05GI92G from the Beyond Einstein Foundation Science Program, grant NAG5-13207 from the Origins of Solar Systems Program, and grant NNG04GL01G from the Astrophysics Theory Program, and by the NSF under grants AST--0307502 and AST--0407040.


\appendix
\section{Wind profiles}\label{sec:wind_profs}
The radial mass-loss profiles of the diffuse- and direct-field photoevaporative winds are defined as follows.  In the ``diffuse'' regime, where the inner disc is optically thick to ionizing photons, the wind is driven by recombinations in the disc atmosphere.  We adopt the wind profile from the hydrodynamic simulations of \citet[][kindly provided in numerical form by Ian McCarthy]{font04}:
\begin{equation}
\dot{\Sigma}_{\mathrm {diffuse}}(R) = 2 n_0(R) u_{\mathrm {l}}(R) \mu m_{\mathrm H} \, .
\end{equation}
Here $n_0(R)$ is the density at the base of the ionized layer, $u_{\mathrm {l}}(R)$ the wind launch velocity and $\mu$ the mean molecular weight of the ionized gas (taken to be $\mu=1.35$ throughout).  The base density profile is given by the fitting form
\begin{equation}
n_0(R)=n_{\mathrm g} \left[ \frac{2}{\left(\frac{R}{R_{\mathrm g}}\right)^{15/2} + \left(\frac{R}{R_{\mathrm g}}\right)^{25/2}}\right]^{1/5} \, ,
\end{equation}
where the base density at $R_{\mathrm g}=GM_*/c_{\mathrm s}^2$ (where $c_{\mathrm s} = 10$km s$^{-1}$ is the sound speed of the ionized gas) is given by
\begin{equation}
n_{\mathrm g} = C_1\left(\frac{3 \Phi}{4 \pi \alpha_{\mathrm B} R_{\mathrm g}^3}\right)^{1/2} \, .
\end{equation}
Here $\Phi$ is the stellar ionizing flux (in units of photons s$^{-1}$) and $\alpha_{\mathrm B}$ is the Case B recombination coefficient for atomic hydrogen at $10^4$K, which has a value of $\alpha_{\mathrm B}=2.6\times10^{-13}$cm$^3$s$^{-1}$ \citep{allen}, and the constant $C_1 \simeq 0.14$.  The numerical simulations show that the launch velocity profile is well-matched by the numerical form
\begin{equation}
u_{\mathrm l}(R) = c_{\mathrm s}\, A \exp\left[B \left(\frac{R}{R_{\mathrm g}} -0.1\right)\right] \left(\frac{R}{R_{\mathrm g}} -0.1\right)^D \, , \quad R\ge0.1R_{\mathrm g}
\end{equation}
where the numerical constants have the values $A=0.3423$, $B=-0.3612$ and $D=0.2457$.  For radii smaller than $0.1R_{\mathrm g}$ the wind rate is set to zero.

In the ``direct'' regime, where the inner disc is optically thin to ionizing photons, the wind is instead driven by direct illumination of the inner disc edge.  In this case we adopt the wind profile from the simulations of \citet{acp06a}.  The base density at the inner disc edge is set by ionization balance (essentially a Str\"omgren criterion), and the hydrodynamic simulations show that the following power-law form provides a good fit to the wind profile:
\begin{eqnarray}
\dot{\Sigma}_{\mathrm {direct}}(R,t) = 2 C_2 \mu m_{\mathrm H} c_{\mathrm s}  \left(\frac{\Phi}{4\pi \alpha_{\mathrm B} (H/R) R_{\mathrm {in}}^3(t)}\right)^{1/2} 
\\
\nonumber \times \left(\frac{R}{R_{\mathrm {in}}(t)}\right)^{-a}\, , \quad R>R_{\mathrm {in}} \, .
\end{eqnarray}
Here $R_{\mathrm {in}}(t)$ is radius of the inner disc edge, which increases as the wind disperses the outer disc.  The fitting constants $C_2$\footnote{Note that the constant defined as $C_2$ here is denoted by ``$CD$'' in \citet{acp06a}.} and $a$ vary somewhat with the disc aspect ratio $H/R$, but for simplicity we adopt the values derived for $H/R=0.05$: $C=0.235$ and $a=2.42$.

\label{lastpage}

\end{document}